\documentclass[aps,pra,reprint,10pt,twocolumn,superscriptaddress,floatfix]{revtex4-1}
\usepackage{times}
\usepackage{amsmath,amssymb,bm,graphicx}
\usepackage[breaklinks,colorlinks=true,urlcolor=blue,citecolor=blue,linkcolor=blue]{hyperref}
\usepackage{mathtools}
\usepackage{xcolor}
\usepackage{enumitem}
\usepackage{braket}
\usepackage{xspace}
\usepackage{grffile}
\usepackage[caption=false]{subfig}

\newcommand{\figfac}{1.0}

\begin{document}

\title{Efficient matrix-product-state preparation of highly entangled trial states: \\ Weak Mott insulators on the triangular lattice revisited}

\date{\today}

\author{Amir M Aghaei}
\affiliation{Department of Physics and Astronomy, University of California at Riverside, Riverside, California 92521, USA}

\author{Bela Bauer}
\affiliation{Microsoft Station Q, University of California, Santa Barbara, California 93106-6105, USA}

\author{Kirill Shtengel}
\affiliation{Department of Physics and Astronomy, University of California at Riverside, Riverside, California 92521, USA}

\author{Ryan V. Mishmash}
\affiliation{Department of Physics, University of California, Berkeley, California 94720, USA}
\affiliation{Department of Physics, Princeton University, Princeton, New Jersey 08540, USA}

\begin{abstract}
Using tensor network states to unravel the physics of quantum spin liquids in minimal, yet generic microscopic spin or electronic models remains notoriously challenging. A prominent open question concerns the nature of the insulating ground state of two-dimensional half-filled Hubbard-type models on the triangular lattice in the vicinity of the Mott metal-insulator transition, a regime which can be approximated microscopically by a spin-1/2 Heisenberg model supplemented with additional ``ring-exchange'' interactions. Using a novel and efficient \emph{state preparation} technique whereby we initialize full density matrix renormalization group (DMRG) calculations with highly entangled Gutzwiller-projected Fermi surface trial wave functions, we show---contrary to previous works---that the simplest triangular lattice $J$-$K$ spin model with four-site ring exchange likely does \emph{not} harbor a fully gapless U(1) spinon Fermi surface (spin Bose metal) phase on four- and six-leg wide ladders. Our methodology paves the way to fully resolve with DMRG other controversial problems in the fields of frustrated quantum magnetism and strongly correlated electrons.
\end{abstract}

\maketitle

{\bf \emph{Introduction.}}~Quantum spin liquids (QSLs) are elusive states of quantum matter that defy usual ordering down to very low temperatures, contain long-range quantum entanglement, and exhibit nontrivial quasiparticle excitations~\cite{balents_spin_2010, savary_quantum_2017, zhou_quantum_2017, broholm_quantum_2020}. Such behavior is often caused by frustration, which makes the ground state (and its low-energy excitations) a system-wide compromise between extensively many quantum degrees of freedom~\cite{Anderson73, Anderson87}. %The resonating valence bond (RVB) state initially proposed by Anderson~ was the first example of a QSL. %Such gapped spin liquids have since been studied for their fundamentally interesting behavior as well as for their potential application to robustly store quantum information and perform fault-tolerant quantum computation~\cite{kitaev_fault-tolerant_2003, kitaev_anyons_2006}.
Despite their long-range entanglement, QSLs with a finite correlation length are relatively tractable to study with tensor-network-state simulations~\cite{orus_practical_2014}, at least in two spatial dimensions (2D)~\cite{PhysRevLett.110.067208, PhysRevB.91.224431, PhysRevB.89.075110}. The general understanding of \emph{gapless} QSLs in 2D is however particularly limited. This is in part because the study of gapless phases has long been hampered by the inability of numerical tools to catch up with theory and even experiment. With a diverging correlation length, typical simulations usually require a large number of spins to reliably identify the nature of the state, and exact diagonalization methods may at best be able to suggest a possible lack of ordering~\cite{misguich_spin-liquid_1999, liming_nelong-range_2000}.

Certain gapless spin liquids are characterized by \emph{emergent Fermi surfaces}, thereby leading to a multiplicative log correction to the usual boundary law of entanglement entropy~\cite{zhang_entanglement_2011, mishmash_entanglement_2016, hu_fractionalized_2020}. This property renders such states particularly challenging to attack with density matrix renormalization group (DMRG)---still the gold standard tensor-network-based technique in the field~\cite{stoudenmire_studying_2012}---when approaching 2D: For a $L_y \times L_x$ system with $L_x \gg L_y$, the entanglement grows as $S \sim L_y \log L_x$, implying that the required matrix product state (MPS) bond dimension scales as a daunting $M \sim e^S \sim (A\,L_x^\alpha)^{L_y}$ (for some constants $A > 1$ and $\alpha \geq 0$~\footnote{In the case of a gapped spin liquid, one would expect $\alpha=0$.}). We hereafter refer to such states as \emph{highly entangled}.

\begin{figure}[b]
  \centering
  \includegraphics[width=\figfac\columnwidth]{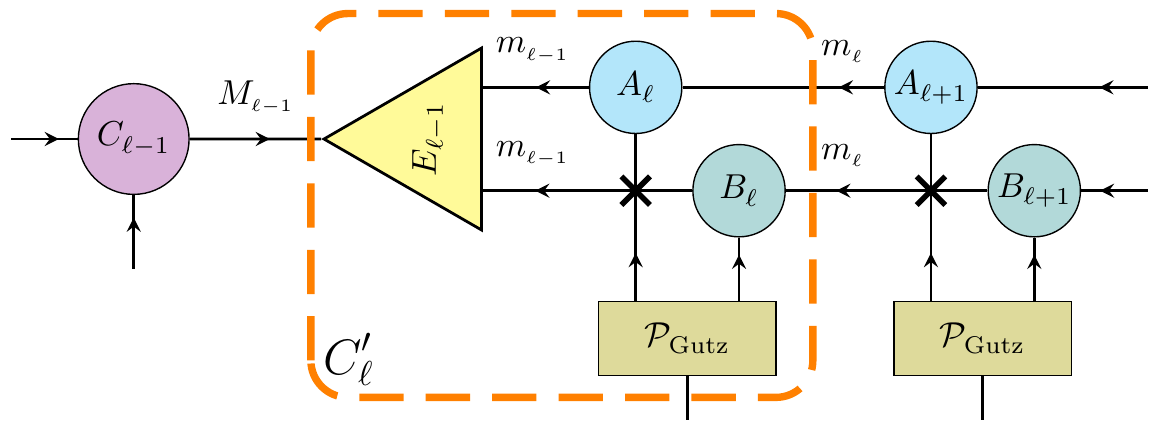}
  \caption{Illustration of a single step of the \emph{Gutzwiller zipper} method. During this process, two MPSs (cyan and green) representing single-component fermionic Gaussian states are ``zipped'' (see text) to give a single MPS (purple) for the final spin-$1/2$ Gutzwiller-projected wave function.
  \label{fig:zipandgutz}
  }
\end{figure}

{\bf \emph{Spinon Fermi surface state and ring-exchange model.}}~The most prominent example of a highly entangled gapless QSL with emergent Fermi surfaces is the U(1) spinon Fermi surface (SFS) state \cite{lee_u1_2005, motrunich_variational_2005} (also known as the ``spin Bose metal''~\cite{sheng_spin_2009, block_spin_2011}). Its low-energy description involves decomposing the physical spin operator $\mathbf{S}_i$ in terms of spin-$1/2$ fermions, \emph{spinons}, at half-filling (one spinon per lattice site), taking the spinons to form a gapless Fermi surface state at the mean-field level, and finally coupling the spinons to an emergent U(1) gauge field. At the level of a variational wave function, the essential physics of the U(1) SFS field theory can be captured by performing a simple \emph{Gutzwiller projection} on the mean-field state~\footnote{For some potential caveats, see Ref.~\cite{tay_failure_2011}.}. The resulting Gutzwiller-projected Fermi surface (GPFS) wave function reads:
\begin{equation}
    |\Psi_\mathrm{GPFS}\rangle = \mathcal{P}_\mathrm{Gutz} |\Psi_\mathrm{MF}\rangle,
    \label{eq:GPFS}
\end{equation}
where $|\Psi_\mathrm{MF}\rangle$ is the mean-field Gaussian fermionic state of spinons (forming a Fermi surface), and
$\mathcal{P}_\mathrm{Gutz} \equiv \prod_i (n_{i \uparrow} - n_{i\downarrow})^2$
projects out all components of the fermionic wave function with doubly-occupied or empty sites. %, thereby recovering a wave function in the Hilbert space of spins-$1/2$.
%This remarkable wave function inherits many properties of the fermionic mean-field state---in particular ``$2k_F$'' singularities in the spin structure factor~\cite{motrunich_variational_2005}, nearly the same entanglement entropy scaling~\cite{zhang_entanglement_2011}, etc.---but it is after all a wave function for spins-$1/2$. %(i.e., a ``perfect insulator'').

Unconventional as it may seem, this remarkable U(1) spinon Fermi surface state is actually a very natural theoretical description for \emph{weak Mott insulators} \cite{lee_u1_2005, motrunich_variational_2005, senthil_theory_2008, lai_two-band_2010, mishmash_continuous_2015} %---i.e., electrons just on the insulating side of the (bandwidth-driven) Mott metal-insulator transition (assumed to occur at finite interaction strength \footnote{If the Fermi surface is nested (e.g., the nearest-neighbor hopping square lattice) such that the Mott transition occurs at $U = 0^+$, the projected wave function, Eq.~\eqref{eq:GPFS}, actually magnetically orders~\cite{li_antiferromagnetic_2011, zhang_entanglement_2011}.})---
and is thus a strong candidate for the low-energy description of real materials believed to be in this regime~\cite{kanoda_mott_2011}. 
% To capture the microscopics of weak Mott insulators at the level of a spin Hamiltonian (as opposed to a full electronic model), one can perform a strong-coupling expansion in the inverse interaction strength (e.g., $1/U$)~\cite{macdonald_fractu_1988, yang_effective_2010}. Specializing to the 2D triangular lattice for concreteness, for the Hubbard model, the leading term is the \ks{drop: familiar SU(2)-invariant} nearest-neighbor Heisenberg exchange ($J \sim t^2/U$), while the next term is a 4-site cyclic ``ring-exchange'' interaction ($K \sim t^4/U^3$). A minimalist spin Hamiltonian to explore is then the ``$J$-$K$'' model~\cite{motrunich_variational_2005, block_spin_2011, grover_weak_2010, mishmash_theory_2013, he_spinon_2018}:
Specializing to the 2D triangular lattice, a popular minimal spin Hamiltonian to describe the physics of weak Mott insulators is the ``$J$-$K$'' ring-exchange model~\cite{motrunich_variational_2005, block_spin_2011, grover_weak_2010, mishmash_theory_2013, he_spinon_2018, macdonald_fractu_1988, yang_effective_2010}:
\begin{equation}
  H = \sum_{\langle i,j\rangle} 2J_{ij}\,\mathbf{S}_i\cdot \mathbf{S}_j
  + \sum_{ijkl \in \Diamond} K_{\Diamond}\,(P_{ijkl} + \mathrm{H.c.}),
  \label{eq:JKmodel}
\end{equation}
where we take isotropic couplings ($J_{ij} = J, K_\Diamond = K$) unless otherwise noted, and $J$ and $K$ are taken to vary independently. %(i.e., not tied to any underlying electronic model).

There is in fact tantalizing evidence suggesting that the U(1) SFS may in fact be the correct low-energy description of Eq.~\eqref{eq:JKmodel}, at least for sufficiently large $K/J \gtrsim 0.3$. Firstly, the GPFS trial wave function [Eq.~\eqref{eq:GPFS}] has remarkably favorable ring-exchange energy, thus making it the best variational state found to date in this parameter regime~\cite{motrunich_variational_2005, grover_weak_2010, mishmash_theory_2013}. Furthermore, a series of DMRG studies on 2-leg~\cite{sheng_spin_2009}, 4-leg~\cite{block_spin_2011}, and 6-leg~\cite{he_spinon_2018} wide ladder geometries similarly points to a stable SFS phase. As emphasized above, the U(1) SFS represents a highly entangled ground state; specifically, when placed on a quasi-1D cylindrical geometry, the bipartite entanglement entropy (for $\ell$-site subsystems, using the usual DMRG ``snake'' path, embedded in $N = L_y \times L_x$ total sites) scales as~\cite{0405152}
\begin{equation}
    S_1(\ell, N = L_y \times L_x)
    = \frac{c}{6}\log\left(\frac{N}{\pi}\sin\frac{\pi\ell}{N}\right) + A',
    \label{eq:cardy}
\end{equation}
where the effective central charge $c = 2N_\mathrm{slices} - 1$ with $N_\mathrm{slices} \sim L_y$ the number of ``slices'' through which the quantized tranverse momenta pierce the emergent Fermi surface (i.e., the number of \emph{partially filled} 1D spinful bands at the mean-field level)~\cite{sheng_spin_2009, geraedts_half-filled_2016}. However, only the 2-leg study~\cite{sheng_spin_2009} of the $J$-$K$ spin model was able to conclusively confirm that the DMRG ground state has the expected $c = 3$ (for $N_\mathrm{slices} = 2$). On the other hand, the 4-leg~\cite{block_spin_2011} and 6-leg~\cite{he_spinon_2018} studies reached their conclusions mainly based on analysis of equal-time correlation functions, but were unable to pin down the expected central charges of $c = 5$ ($N_\mathrm{slices} = 3$) and $c = 9$ ($N_\mathrm{slices} = 5$), respectively. At this point, it is not clear if the issue is entirely due to insufficient number of DMRG states kept (MPS bond dimension), i.e., lack of convergence, or if there is physics at play: Perhaps the true ground state exhibits an instability of the U(1) SFS and the true $c < 2N_\mathrm{slices} - 1$?

{\bf \emph{State preparation strategy.}}~We develop a scheme capable of addressing this ambiguity directly by focusing on \emph{state preparation}, i.e., initialization of the DMRG energy optimization procedure. The importance of deliberate initial state preparation is ubiquitous in many areas, from classical optimization problems~\cite{gondzio_warm_1998} to variational~\cite{mcclean_barren_2018, zhou_quantum_2020, egger_warm-starting_2020} and fault-tolerant~\cite{reiher_elucidating_2017, tubman_postponing_2018} quantum algorithms---we here illustrate its utility in the context of DMRG. In particular, we devise an efficient means to construct a faithful finite-size MPS representation of the (highly entangled) GPFS trial wave function [Eq.~\eqref{eq:GPFS}] via a significant improvement of the approach first proposed in Ref.~\cite{bauer2019symmetry} (see Fig.~\ref{fig:zipandgutz}). We then ``warm start'' the DMRG optimization using this GPFS MPS as the initial state.

If we can accurately represent the trial state as an MPS with a given bond dimension $M$ and capture its expected entanglement entropy scaling ($c = 2N_\mathrm{slices} - 1$), then it is natural to expect that we can capture the entanglement of the DMRG ground state itself---whether the latter in fact realizes the U(1) SFS or some instability thereof, which will in general have lower entanglement~\footnote{The U(1) SFS can be thought of as a ``mother'' state~\cite{hermele_algebraic_2005}; furthermore, it is the most highly entangled plausible universality class of the model.}. If the true DMRG ground state is in fact in the same universality class, the DMRG iteration will only change short-range properties of the state. In most systems, the contributions of such short-range correlations to the entanglement are small compared to the universal contributions from the gapless modes~\footnote{Although some counterexamples exist; see, e.g., Ref.~\cite{bauer2019symmetry}.}, making it possible for the true ground state to be captured with comparable bond dimension. If, on the other hand, the true ground state corresponds to an instability of the trial state with lower entanglement entropy, we expect DMRG energy optimization to \emph{decrease} the entanglement. In our case, such behavior would provide strong evidence \emph{against} the hypothesis that the DMRG ground state realizes the U(1) SFS. %Finally, it is possible for the true ground state to lie in a different, more entangled universality class, in which case DMRG optimization of the trial state may not converge at comparable bond dimension. In our case, there is no plausible candidate for such an even more highly entangled universality class, and this scenario thus seems unlikely.
Below, we benchmark and apply this philosophy to the problem of the $L_y = 2, 4$, and 6 leg wide $J$-$K$ ring-exchange model introduced above.

{\bf \emph{The Gutzwiller zipper.}}
To efficiently construct an MPS represenation of the GPFS wave function, we first use the prescription of Fishman and White~\cite{fishman_compression_2015} to build as MPSs two identical~\footnote{For the SU(2) invariant U(1) SFS ansatz, the mean-field hopping parameters for the two species of spinons are identical.} fermionic Gaussian states through a series of $\mathcal{O}(N)$ Givens rotations. This approach is basically identical to that used for preparing Slater determinant states in an arbitrary basis on a quantum computer~\cite{wecker_solving_2015, kivlichan_quantum_2018}, as implemented recently in quantum hardware~\cite{collaborators_hartree-fock_2020}. We obtain a ``parton MPS'' for the $\uparrow$ spinons given by $\ket{\psi_\uparrow} = \sum_{\vec{n}_\uparrow} A_1^{n_{\uparrow 1}} \ldots A_N^{n_{\uparrow N}} |\vec{n}_\uparrow\rangle$,
% \begin{equation} \label{eqn:mps-definition}
% \ket{\psi_\uparrow} = \sum_{\vec{n}_\uparrow} A_1^{n_{\uparrow 1}} \ldots A_N^{n_{\uparrow N}} |\vec{n}_\uparrow\rangle,
% \end{equation}
where the $\vec{n}_\uparrow$ are occupation-number vectors and each $A_\ell^{n_{\uparrow\ell}}$ is a matrix of size $m_{\ell-1} \times m_\ell$; $m_\ell$ is the so-called  bond dimension. Likewise, we denote the matrices that form the MPS for the $\downarrow$ spinons by $B_\ell$.
% We denote the obtained ``parton'' MPSs for the $\uparrow$ and $\downarrow$ spinons as $A_\ell$ and $B_\ell$ with bond dimensions $m_\ell$ at bond $\ell$ and physical (local) dimension $d=2$.
The only discernible error incurred thus far is the truncation to $m_\ell$ states.

To proceed, one could naively form the tensor product state $\ket{\Psi_\mathrm{MF}} = \ket{\psi_\uparrow} \otimes \ket{\psi_\downarrow}$ and then perform the Gutzwiller projection. However, the bond dimension of the tensor product state will be the product of the bond dimensions of each constituent state, and this procedure will thus scale as $\mathcal{O}(m_\ell^6)$. %~\footnote{For simplicity, we assume throughout this discussion that the MPS bond dimension for each spin species is equal.}.
On the other hand, since the Gutzwiller projection reduces the entanglement entropy of the state, one may expect that the bond dimension required to accurately describe $\ket{\Psi_\mathrm{GPFS}}$ is much smaller than that required for the tensor product state $\ket{\Psi_\mathrm{MF}}$.
To overcome this issue, we perform the tensor product, Gutzwiller projection, and truncation to a new MPS of bond dimension $M_\ell \ll m_\ell^2$ on each bond in one iterative sweep, which we refer to as the ``\emph{Gutzwiller zipper}''.
% Next, we perform the tensor product and Gutzwiller projection together in an iterative fashion using a highly optimized scheme that we refer to as the ``\emph{Gutzwiller zipper}''.

Assuming that the $A_\ell$ and $B_\ell$ MPSs are in canonical form with orthogonality center at the first site, we perform the following steps for all sites (see Fig.~\ref{fig:zipandgutz}): (\emph{i}, ``zip'') Form the matrix $C'_{\ell}$ as the tensor contraction shown in the orange, dashed box in Fig.~\ref{fig:zipandgutz} comprising $A_{\ell}, B_{\ell}$, the Gutzwiller projection operator, and the carry from the previous step $E_{\ell-1}$ ($E_0 \coloneqq I$). (\emph{ii}, truncate) Bundle the physical dimension with the left index of $C'_{\ell}$, perform a singular value decomposition (SVD) as $C'_{\ell} = U_{\ell} S_{\ell} V^{\dagger}_{\ell}$, and truncate to $M_\ell$ singular values. (\emph{iii}) Identify $U_{\ell}$ as the MPS tensor corresponding to the truncated tensor product MPS at site $\ell$: $C_{\ell} \coloneqq U_{\ell}$, and identify $S_{\ell}V^{\dagger}_{\ell}$ as the carry matrix for the next step: $E_{\ell} \coloneqq S_{\ell}V^{\dagger}_{\ell}$. In the last step, $C_L \coloneqq C'_L$. The resulting MPS
% is an approximation to the Gutzwiller projection of the full tensor product state; it
has bond dimensions $M_\ell$ and is in canonical form with its orthogonality center at the last site $N$. One full sweep takes $\mathcal{O}(Nm^2M^2)$ operations. %(see~\cite{suppmat} for details).
While for the expected regime $m \ll M \ll m^2$, one might expect the Gutzwiller zipper to take longer than a DMRG sweep on the resulting MPS, we in practice find that for the relevant parameters used below, the prefactor of the zipper is much lower and it is in fact computationally cheaper than the subsequent DMRG sweeps. Details of the implementation as well as the correct treatment of the fermionic exchange sign can be found in~\cite{suppmat}. (For alternative approaches, see Refs.~\cite{wu_tensor_2020, jin_efficient_2020}.)

\begin{figure}[t]
  \centering
  \includegraphics[width=\figfac\columnwidth]{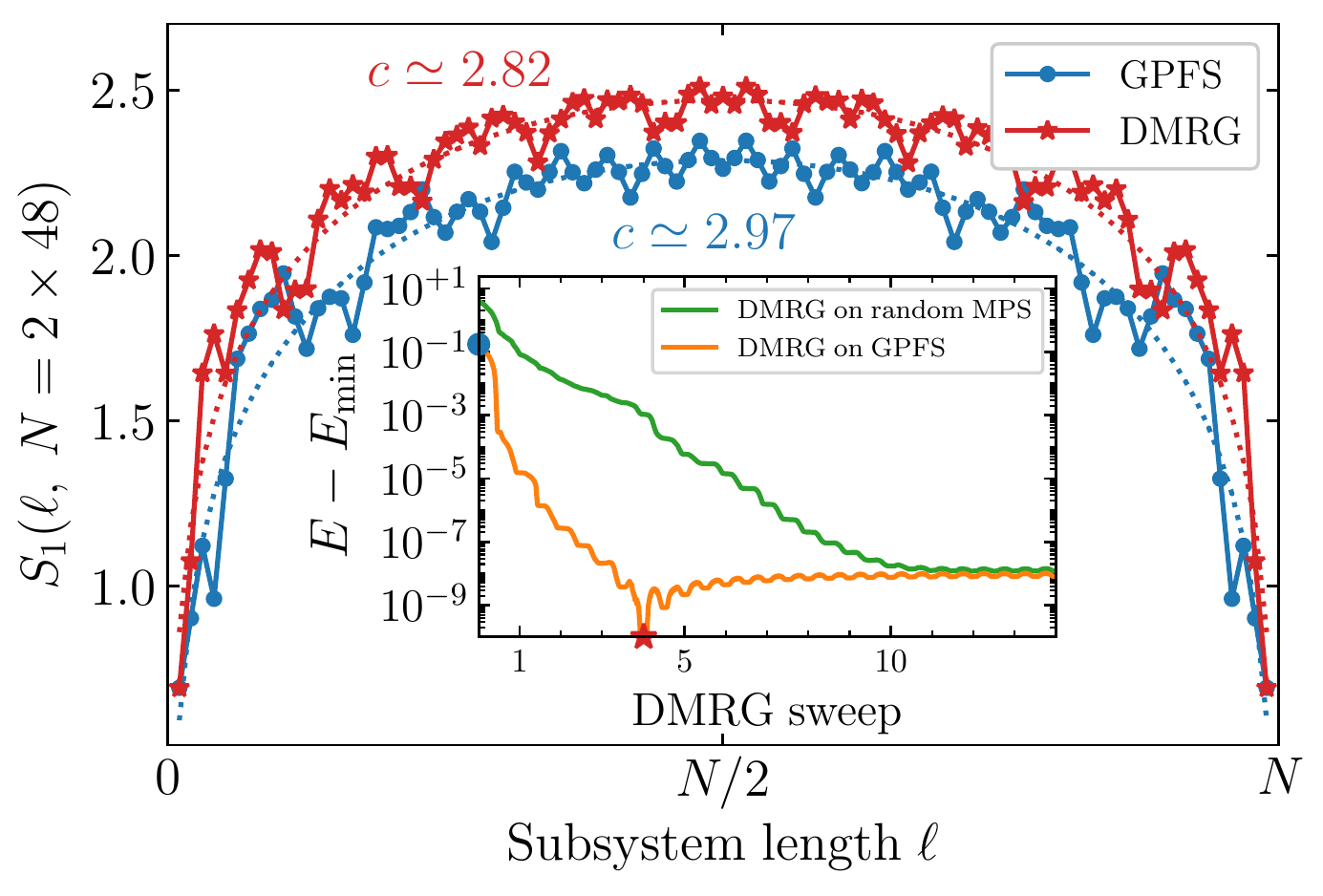}
  \caption{Entanglement entropy $S_1$ versus subsystem length $\ell$ on the 2-leg triangular strip for the Gutwiller-zipper-obtained GPFS MPS and final DMRG ground state ($N=2\times48$). In the inset, we show the DMRG energy per site during sweeping for both initialization procedures (the points marked on the orange curve correspond to the respective data in the main panel), where $E_\text{min}$ is the minimum value achieved during the DMRG process (with GPFS initialization). %(entanglement and correlation functions).
  \label{fig:2legcomparison}
  }
\end{figure}

{\bf \emph{Fate of the SFS in the triangular lattice $J$-$K$ model.}}~We begin by benchmarking our approach on the 2-leg triangular strip $J$-$K$ model. This model was solved originally in an extensive study by Sheng et al.~\cite{sheng_spin_2009} which left essentially zero doubt that the 2-band ($N_\mathrm{slices}=2$) U(1) SFS state is realized in a wide swath of the phase diagram. In particular, a central charge $c = 2N_\mathrm{slices}-1=3$ was confirmed by performing traditional DMRG calculations on a system with periodic boundary conditions (see Figs.~9 and 10 of Ref.~\cite{sheng_spin_2009}). In Fig.~\ref{fig:2legcomparison}, we perform analogous calculations initializing the DMRG with a 2-band GPFS trial state. Here, we choose for the Hamiltonian couplings $J_2/J_1=0.8$ and $K/J_1=1$ (following the conventions of Ref.~\cite{sheng_spin_2009}) and work on a system with open boundary conditions (OBC). For the initial trial state, to generate $|\Psi_\mathrm{MF}\rangle$ we take a mean-field spinon hopping Hamiltonian with $t_2/t_1=0.7$~\cite{suppmat} which gives a generic 2-band parton band-filling configuration for this region of the phase diagram.
%, which is nearly but not precisely the energetically optimal choice.

The main panel of Fig.~\ref{fig:2legcomparison} shows data for the von Neumann entanglement entropy $S_1$ on an $N = 2 \times 48$ triangular strip for the MPS-constructed GPFS with $m=200$ and $M=900$. A fit to the scaling form Eq.~\eqref{eq:cardy} is consistent with $c=3$, where the Friedel-like oscillations are due to the open boundaries. (Note that in the absence of an MPS representation, it is not otherwise known how to calculate $S_{\alpha<2}$ for such projected wave functions~\cite{zhang_entanglement_2011}.) The final DMRG entanglement entropy data after just four DMRG sweeps (each sweep being a left-to-right + right-to-left traversal of the lattice) is also shown: the DMRG entanglement scaling indeed exhibits $c\approx3$ (for details of fitting see~\cite{suppmat}) albeit with a slightly larger constant $A'$ (whereby we increase the bond dimension during DMRG to $M=2000$). Running DMRG on top of the GPFS state indeed very quickly fixes up the nonuniversal short-distance physics (e.g., details of the parton band fillings on the scale of $2\pi/N$, the cutoff-dependent $A'$ term in $S_1$, etc.) and the system rapidly converges. In the inset of Fig.~\ref{fig:2legcomparison}, we show the energy of the ground state as we sweep the DMRG, comparing traditional random state initialization versus our GPFS MPS seeding strategy; the latter converges drastically quicker. In~\cite{suppmat}, we present more tests and sanity checks of our approach using the well-studied 2-leg $J$-$K$ system as a testbed.

\begin{figure}[t]
  \centering
  \includegraphics[width=\figfac\columnwidth]{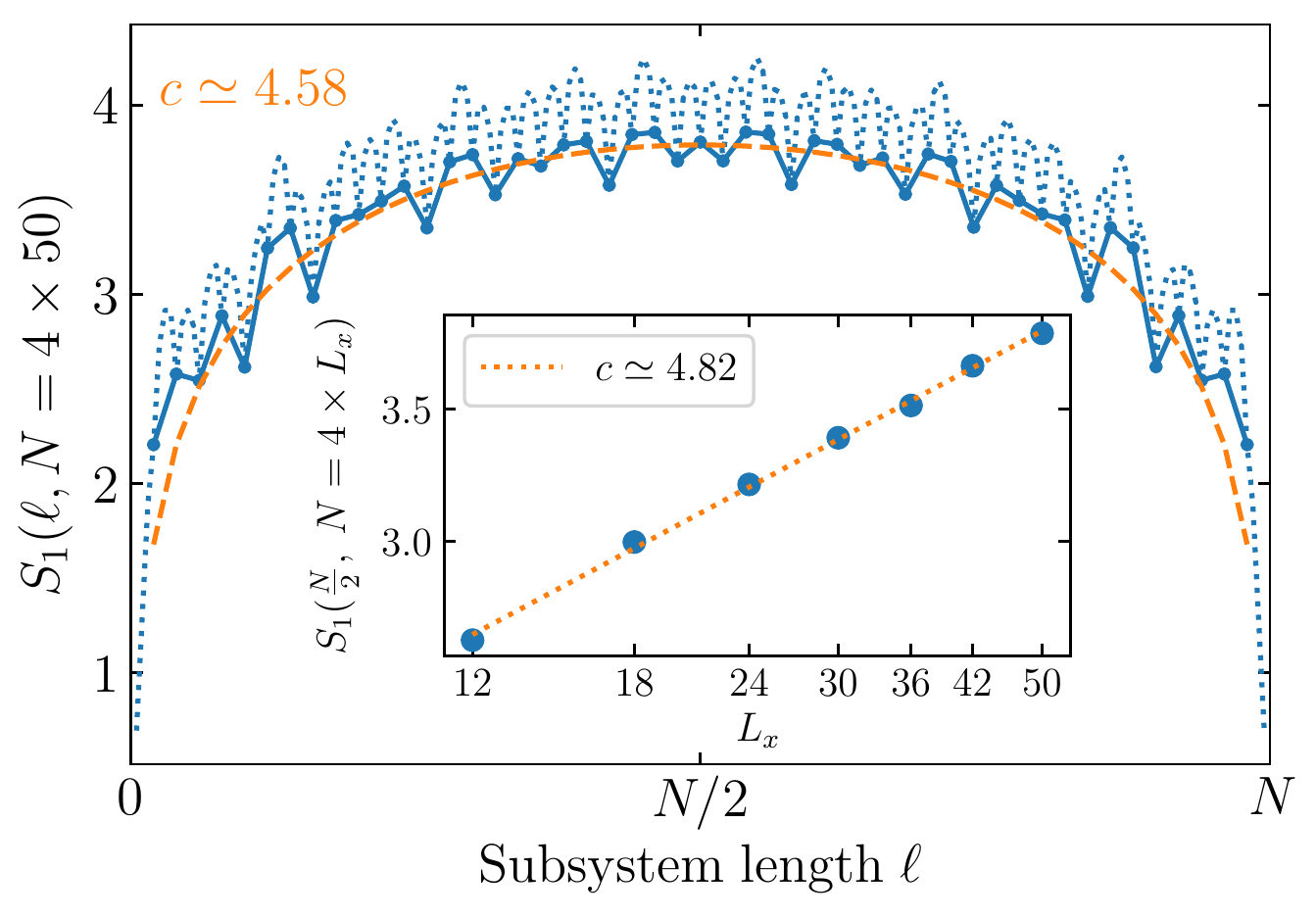}
  \caption{Entanglement entropy for the 3-band GPFS MPS on the 4-leg ladder as obtained by the Gutzwiller zipper method. In the main panel, we take $N=4\times50$ and data connected by solid lines (dotted lines) correspond to ``rung'' cuts (all cuts)~\cite{suppmat} with the dashed orange curve a fit with $c\simeq4.58$. The inset shows $S_1$ evaluated at $\ell=N/2$ for several $N = 4\times L_x$ (linear-log plot), confirming scaling consistent with $c=5$.
  \label{fig:4leggutzvne}
  }
\end{figure}

\begin{figure}[t]
  \begin{center}
  \includegraphics[width=\figfac\columnwidth]{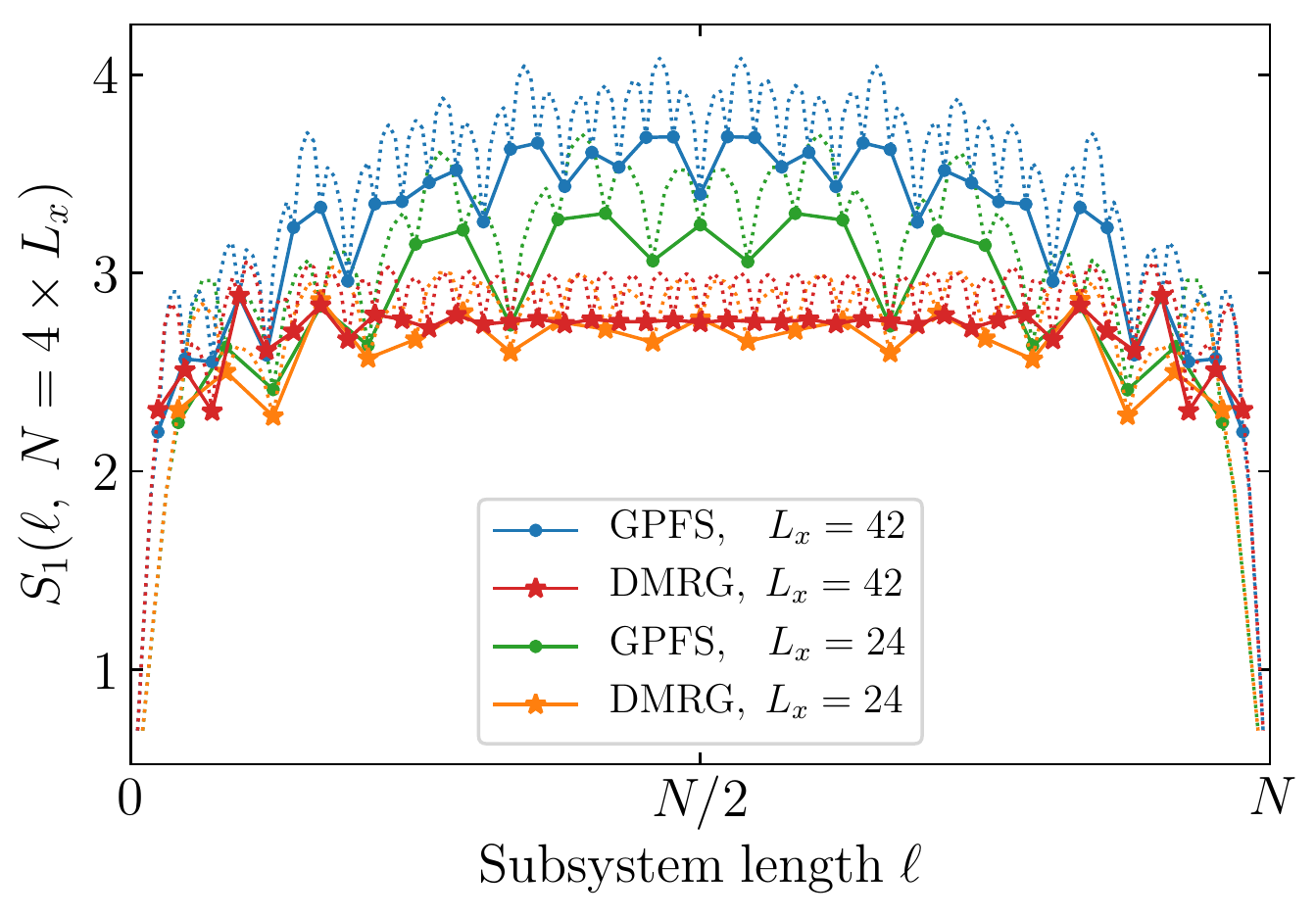}
  \includegraphics[width=\figfac\columnwidth]{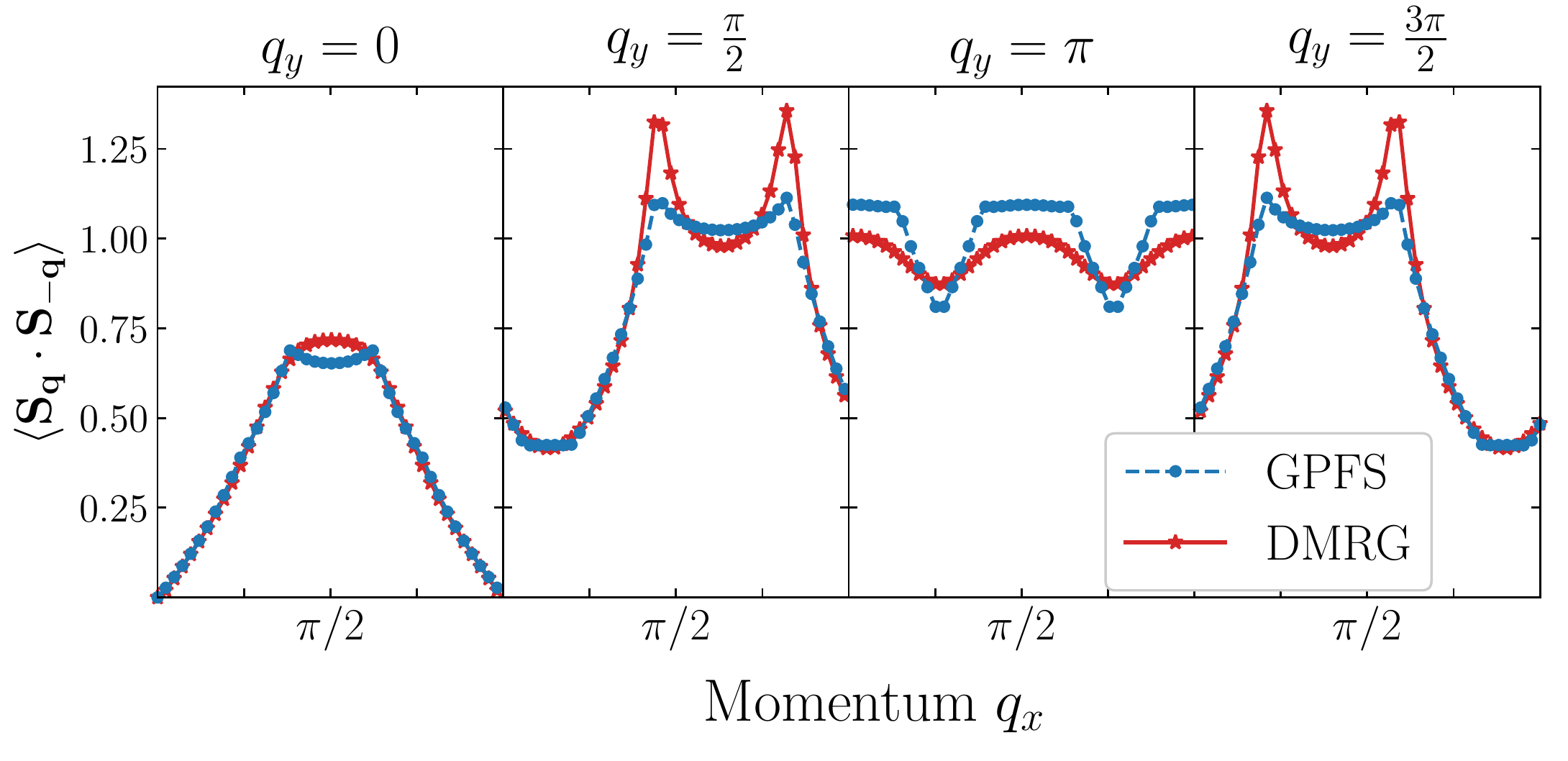}
  \end{center}
  \vspace{-0.2in}
  \caption{Entanglement entropy scaling (top) and spin structure factor (bottom) on the 4-leg ladder for the 3-band GPFS MPS (cf.~Fig.~\ref{fig:4leggutzvne}) and final DMRG ground state of Eq.~\eqref{eq:JKmodel} with $K/J=0.6$.
  \label{fig:4legdmrggutz}
  }
\end{figure}

We now turn to the 4-leg wide $J$-$K$ ladder first studied by Block et al.~\cite{block_spin_2011}. For isotropic Heisenberg and ring-exchange couplings [$J_{ij}=J$ and $K_\Diamond=K$ in Eq.~\eqref{eq:JKmodel}], this work proposed that the 3-band incarnation ($N_\mathrm{slices}=3$) of the U(1) SFS is realized for $K/J \gtrsim 0.3$, a natural extension of the 2-leg results~\cite{sheng_spin_2009} toward 2D. More precisely, it was claimed that the U(1) SFS is at the very least a good \emph{starting point} for understanding the true ground state---this caveat being necessary partly because the expected central charge $c=2N_\mathrm{slices}-1=5$ was not confirmed on large systems. Taking the same line of attack as above, we first calculate the entanglement entropy of the MPS approximation of the GPFS trial state itself (taking isotropic nearest-neighbor spinon hopping parameters $t_{ij}=t$). (Again, we use OBC in the $x$ direction; see~\cite{suppmat} for details of our lattice clusters.) Converging to $c\approx5$ is already somewhat numerically challenging for the 3-band GPFS, but as we show in Fig.~\ref{fig:4leggutzvne} it is indeed possible. Here, we perform large-scale simulations on $4\times L_x$ systems with a series of lengths up to $L_x=50$, taking the bond dimensions as high as $m=500$ and $M=4000$ with corresponding final truncation error $\mathcal{O}(10^{-6})$.

We now assess the fate of the $c=5$ GPFS under DMRG energy optimization at the characteristic putative U(1) SFS point $K/J=0.6$ (cf.~Fig.~5 of Ref.~\cite{block_spin_2011}). Strikingly, after only two DMRG sweeps, the entanglement entropy of the DMRG ground state rapidly \emph{decreases} and almost immediately saturates to very clear $c=0$ behavior; that is, completely \emph{flat} scaling of $S_1$ vs subsystem length $\ell$. In the top panel of Fig.~\ref{fig:4legdmrggutz}, we show the GPFS entanglement entropy and the corresponding DMRG data after two sweeps \footnote{The data is basically invariant under further sweeps. For $N=4\times24$, we have also checked that random state initialization gives the same final $S_1$ values after many sweeps.} %; see~\cite{suppmat} for more discussion of convergence with sweeps and $M$.}
for $4\times L_x$ systems with $L_x=24$ and $42$ taking up to $M=3000$. The bottom panel of Fig.~\ref{fig:4legdmrggutz} depicts the $L_x=42$ spin structure factor $\langle \mathbf{S}_\mathbf{q} \cdot \mathbf{S}_{-\mathbf{q}}\rangle$ for both the GPFS MPS trial state and the obtained final DMRG ground state (for conventions used, see~\cite{suppmat} and Ref.~\cite{block_spin_2011}). Both the GPFS and DMRG results are consistent with the top panel of Fig.~5 in Block et al.~\cite{block_spin_2011}, with the minor differences attributable to different lattice conventions~\cite{suppmat} and boundary conditions (cylindrical in our simulations versus fully periodic in Ref.~\cite{block_spin_2011}). We have also checked that the spin structure factor for the GPFS MPS matches exactly that obtained via a traditional variational Monte Carlo~\cite{gros_physics_1989} evaluation on the same trial state. We thus conclude that the DMRG ground state obtained here and in Ref.~\cite{block_spin_2011} is actually likely fully gapped. While there may be some subtle signs of eventual gap formation in the structure factor data (e.g., some smoothed singularities and a slight drop in the slope of $\langle \mathbf{S}_\mathbf{q} \cdot \mathbf{S}_{-\mathbf{q}}\rangle$ near the $\Gamma$ point \cite{mishmash_continuous_2015}), we find this result quite surprising.

\begin{figure}[t]
  \centering
  \includegraphics[width=\figfac\columnwidth]{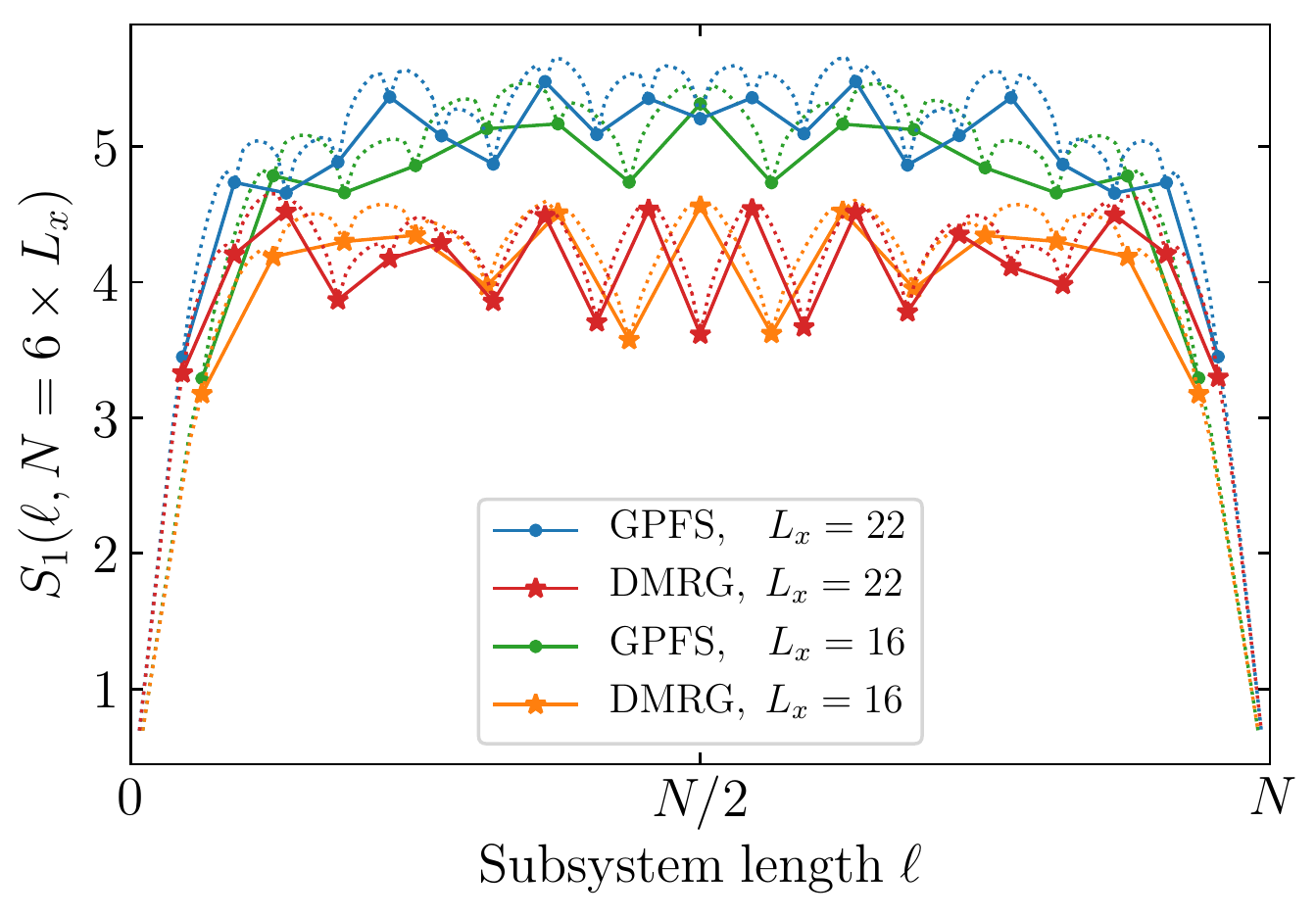}
  \caption{Entanglement entropy scaling of the GPFS MPS (which is not fully converged~\cite{suppmat}) and final DMRG ground state of the $J$-$K$ model (with $K/J=0.6$) on the 6-leg ladder.
  \label{fig:6legdmrggutz}
  }
\end{figure}

Finally, we turn to the 6-leg ladder which may harbor a 5-band ($N_\mathrm{slices}=5$) U(1) SFS~\cite{he_spinon_2018}. In this case, we cannot fully converge the trial state to $c=2N_\mathrm{slices}-1=9$, which we estimate would require excessively large $m\gtrsim 2000$ and $M\gtrsim 10000$ for system sizes considered here. Still, we have constructed an approximate GPFS state via the Gutzwiller zipper using up to $m=1200$ and $M=6000$ on $6\times L_x$ clusters to up length $L_x=22$~\cite{suppmat}, and we expect this MPS to capture short-distance features and sign structure of the phase reasonably well~\cite{suppmat}. Furthermore, the obtained $S_1$ values for the GPFS MPS near the center of the sample are at values significantly above those of Fig.~S11 in Ref.~\cite{he_spinon_2018}~\footnote{Note, however, that the models are not exactly the same; see discussion in Sec.~VIII of the Supplemental Material of Ref.~\cite{he_spinon_2018}.}. The results obtained upon performing subsequent DMRG optimization are shown in Fig.~\ref{fig:6legdmrggutz}. Once again, $\mathcal{O}(1)$ DMRG sweeps quickly \emph{decreases} the entanglement entropy relative to the initial state to a nearly constant scaling versus $\ell$ (modulo quite strong rung-to-rung oscillations), pointing again to a possible $c=0$ state---at the least making an eventual $c=9$ (or $c=8$ for a $Z_2$ SFS~\cite{he_spinon_2018}) result seem unlikely.
%(although perhaps not yet definitively ruled out).

{\bf \emph{Discussion.}}~While we have presented evidence of a possible instability of the U(1) SFS in the $J$-$K$ model on 4- and 6-leg ladders, more work is needed to fully characterize the putative gapped spin liquid state, a task most conveniently done on the infinite cylinder (see Ref.~\cite{petrica_finite_2020} for a very recent implementation of Gutzwiller-projected states as iMPS). In particular, it is interesting to explore connections of our results to the chiral spin liquid state recently observed in the half-filled triangular lattice Hubbard model itself~\cite{szasz_chiral_2020} (cf.~Ref.~\cite{shirakawa_ground-state_2017}). Our results could also be relevant to the recent finding of pair-density-wave superconducting correlations in the doped 4-leg $J$-$K$ ring model~\cite{xu_pair_2019}. Furthermore, we believe our trial wave function state preparation strategy can be robustly used to critically (re-)assess with DMRG prior~\footnote{See, e.g., Refs.~\cite{motrunich_d_2007, sheng_strong-coupling_2008, block_exotic_2011, mishmash_bose_2011, jiang_non-fermi-liquid_2012, bieri_gapless_2015, gong_chiral_2019, li_spinon_2017, m-aghaei_signatures_2018, pereira2018gapless, bauer2019symmetry, hickey_emergence_2019, patel_magnetic_2019, jiang_field-induced_2019, keselman_emergent_2020}} and future claims of emergent Fermi surfaces in generic microscopic models. Finally, it would be interesting to apply our methodology to other open problems in the field, such as the kagome Heisenberg antiferromagnet \cite{yan_spin-liquid_2011, kolley_phase_2015, he_signatures_2017} and the triangular lattice $J_1$-$J_2$ model \cite{hu_competing_2015, zhu_spin_2015, hu_dirac_2019}. While in these cases the smoking-gun leading entanglement entropy scaling / central charge analysis used above will not apply, converging the relevant trial states as MPSs should be less computationally demanding. 
%, and/or to employ other tensor network state frameworks beyond MPS, such as PEPS and MERA.

\emph{Note added}: After completion of this work, the following preprints appeared on the topic of Gutzwiller projection and matrix product states: \cite{petrica_finite_2020,  baiardo_transcorrelated_2020, jin_density_2020}.

{\bf \emph{Acknowledgements.}} We would like to thank Mike Zaletel and Steve White for useful discussions. A.M.\ and K.S.\ were supported in part by the BSF Grant No.~2016255.

\newpage
\bibliography{./refs}

\begin{thebibliography}{87}
\bibitem [{sup()}]{suppmat}%
  \BibitemOpen
  \href@noop {} {}\bibinfo {note} {For more details of our approach and
  simulations plus additional supporting data, please see the
  Appendices.}\BibitemShut {Stop}%
\end{thebibliography}
\clearpage

\section*{\underline{SUPPLEMENTAL MATERIAL}}

\appendix

\section{Detailed discussion of the Gutzwiller zipper method}
\label{sec:zipgutz}

In this appendix, we will discuss some details of the ``Gutzwiller zipper'' approach introduced in the main text. The GPFS model wave function of the U(1) SFS state is obtained by applying the Gutzwiller projection operator to the tensor product of two single-species parton wave functions for the $\uparrow$ and $\downarrow$ spinons each occupying a set of orbitals $k$ [we assume identical orbitals for each species as appropriate for an SU(2) invariant state]. The final spin state reads
\begin{equation}
  \label{eq:gutzwillerstate}
  \ket{\Psi_\mathrm{GPFS}} = \mathcal{P}_\mathrm{Gutz} \left(\prod_{k}
    d^\dagger_{k,\uparrow} d^\dagger_{k,\downarrow} \ket{\Omega}\right),
\end{equation}
where $d^{\dagger}_{k\sigma}$ is the fermionic creation operator for orbital $k$ and flavor $\sigma$, and $\ket{\Omega}$ is the vacuum of the two flavors. The state in parentheses, i.e., the Fermi surface state at the mean-field level, was denoted $|\Psi_\mathrm{MF}\rangle$ in the main text. Finally, $\mathcal{P}_\mathrm{Gutz}$ denotes the Gutzwiller projection operator:
\begin{equation}
  \label{eq:GutzwillerP}
  \mathcal{P}_\mathrm{Gutz} = \prod_i (n_{i\uparrow} - n_{i\downarrow})^2.
\end{equation}

\subsection{Parton MPS construction}

The individual parton MPSs for the $\uparrow$ and $\downarrow$ spinons are constructed by finding a unitary circuit that creates the Slater determinant in the local basis from an initial product state MPS; application of the circuit is achieved through standard time evolution techniques~\cite{vidal2004efficient}. Different ways to generate such a circuit have been studied in the context of state preparation in the quantum computing~\cite{wecker_solving_2015, kivlichan_quantum_2018} and tensor network~\cite{fishman_compression_2015, hyatt_extracting_2017} communities. The details of the process we use are outlined in Ref.~\cite{fishman_compression_2015}, where the nearest-neighbor unitary operators are found from approximate diagonalization of the correlation matrix $\Lambda _{ij}= \langle c^\dagger_ic_j\rangle$, with the final state referred to as a Gaussian MPS (GMPS). We will discuss the appropriate choice of bond dimension $m$ for each parton state below in Sec.~\ref{sec:error}.

\subsection{Fermion sign and symmetries}

As briefly alluded to in the main text, care must be taken to treat the fermionic sign correctly. To illustrate this issue, we can write the MPS for the $\uparrow$ partons more explicitly as
\begin{equation}
\ket{\psi_\uparrow} = \sum_{\vec{n}_\uparrow} A_1^{n_{\uparrow 1}} \ldots A_N^{n_{\uparrow N}} (c_{\uparrow 1}^\dagger)^{n_{\uparrow 1}} \ldots (c_{\uparrow N}^\dagger)^{n_{\uparrow N}} |\Omega\rangle,
\end{equation}
where the $c^{\dagger}_{\sigma i}$ creates a fermion of flavor $\sigma$ on the $i$th site. Now taking the tensor product $\ket{\psi_\uparrow} \otimes \ket{\psi_\downarrow}$, we arrive at a state with the fermionic creation operators ordered such that all $\uparrow$ creation operators come first, and then all $\downarrow$ creation operators. However, the Gutzwiller projection operator $\mathcal{P}_\mathrm{Gutz}$ acts locally on each site. Thus, to evaluate it, it is necessary to commute the fermion operators such that they are ordered according to physical locality, that is 
\begin{equation}
    (c_{1 \uparrow}^\dagger)^{n_1 \uparrow} (c_{1 \downarrow}^\dagger)^{n_1 \downarrow} \ldots (c_{N \uparrow}^\dagger)^{n_N \uparrow} (c_{N \downarrow}^\dagger)^{n_N \downarrow}
\end{equation}

A convenient way to achieve this in the tensor network language is the fermionic swap tensor introduced in Refs.~\cite{corboz2010mera,corboz2010peps}. This formalism is based on the fact that any fermionic model has at least a $\mathbb{Z}_2$ fermion parity symmetry, which can be implemented on the level of each tensor, i.e. the states on the bond of each tensor can be assigned a label corresponding to even and odd fermionic parity. It can be shown that for a fixed planar representation of the tensor network, the fermionic exchange sign can be taken into account by placing a ``swap tensor'' at each crossing of lines. This tensor is diagonal and evaluates to -1 when the parity on all bonds is odd, and +1 otherwise. It is shown as a black cross in Fig.~\ref{fig:zipandgutz}. In principle, one can rearrange the lines and arrive at a different representation of the tensor network with differently placed swap tensors; this corresponds to a different gauge choice and all physical observables will be identical.

Beyond the fermion parity symmetry, many models exhibit additional symmetries such as U(1) particle number conservation or (pseudo-)spin SU(2) symmetries. These are routinely taken into account in the tensor network representation and lead to computational speedup. In our simulations, we make use of the respective $\uparrow$ and $\downarrow$ charge conservation symmetries generated by $N_\uparrow = \sum_i c_{\uparrow i}^\dagger c_{\uparrow i}$ and $N_\downarrow = \sum_i c_{\downarrow i}^\dagger c_{\downarrow i}$ in the construction of the parton MPS, and the spin U(1) symmetry for the GPFS (which is a subgroup of the full SU(2) spin symmetry, and can be understood as being generated by $N_\uparrow - N_\downarrow$; the sum $N_\uparrow - N_\downarrow$ is trivial after Gutzwiller projection).

\subsection{Scaling of the Gutzwiller zipper}
The method starts with the two MPSs in canonical form with the orthogonality center at the first site and bond dimensions $m_\ell$ at bond $\ell$, and the final output is an approximation to the Gutzwiller projected MPS with bond dimensions $M_\ell$, see Fig.~\ref{fig:zipandgutz}.

In the first step we make the matrix $C'_{\ell}$ that is the tensor contraction of parton MPSs $A_{\ell}, B_{\ell}$ (tensors with dimensions $m_{\ell-1}\times d\times m_{\ell}$), the fermionic swap tensor, the carry from the previous step $E_{\ell-1}$ (a tensor with dimensions $M_{\ell-1}\times m_{\ell-1}\times m_{\ell-1}$), and the Gutzwiller projection tensor. (Here, the physical dimension $d=2$ for spins-1/2.) The best choice of contraction order for these tensors takes
  \begin{equation}
    \mathcal{O}(M_{\ell-1}m_{\ell-1}^2m_\ell d+M_{\ell-1}m_{\ell-1}m_\ell^2d^2)
  \end{equation}
operations or approximately $\mathcal{O}(Mm^3)$. In the next step we bundle the physical dimension and the left index of $C'_{\ell}$ and perform an SVD on the resulting matrix, i.e., $C'_{\ell} = U_{\ell} S_{\ell}V^{\dagger}_{\ell}$, followed by the truncation step that keeps up to $M_\ell$ singular values (states). The SVD is performed on a $dM_{\ell-1}\times m_\ell^2$ matrix, which at best takes
  \begin{equation}
    \mathcal{O}[\min(d^2M_{\ell-1}^2m_\ell^2,dM_{\ell-1}m_\ell^4)]\approx \mathcal{O}(M^2 m^2).
  \end{equation}
This is the most expensive step which sets the general scaling of the method, i.e., $\mathcal{O}(M^2m^2)$. We then use the result of SVD to identify $U_{\ell}$ as the MPS tensor corresponding to the $\ell$th 3-leg tensor of the truncated Gutzwiller-projected MPS, $C_{\ell} \coloneqq U_{\ell}$. We then identify $S_{\ell}V^{\dagger}_{\ell}$ as the carry matrix for the next step: $E_{\ell} \coloneqq S_{\ell}V^{\dagger}_{\ell}$.

The procedure is initialized by defining $E_0 \coloneqq I$ and is terminated by reaching the last step and defining $C_L \coloneqq C'_L$, giving the resulting GPFS MPS in canonical form with center at the last site $N$. Therefore the overall scaling of the Gutzwiller zipper is $\mathcal{O}(NM^2m^2$).

\subsection{Accuracy limitations of the Gutzwiller zipper}
\label{sec:error}

We can think of the tensors $A_1$ through $A_{\ell-1}$ of an MPS as defining an (incomplete) basis for the sites $1$ through $\ell-1$; this basis is enumerated by the right index of $A_{\ell-1}$; when the tensors are canonical, this basis is orthonormal. Likewise, the tensors $A_{\ell+2}$ through $A_N$ define a basis for the sites $\ell+2$ through $N$. When we truncate the bond between $A_\ell$ and $A_{\ell+1}$---for example by contracting the two tensors together, performing an SVD, and then truncating the singular values---we implicitly perform a truncation of the full MPS state. While such an SVD is a locally optimal truncation, its effect on the global state and the accuracy of the approximation clearly depend on the basis defined by the tensors $A_1$ through $A_{\ell-1}$ and $A_{\ell+2}$ through $A_N$.

This leads to an important source of error in the Gutzwiller zipper method. When performing the truncation in step \emph{(ii)} of the Gutzwiller zipper (in a left-to-right sweep), the tensors to the left are in canonical form for a truncated GPFS, while the sites to the right are implicitly in the canonical form for the unprojected tensor product of the two parton MPS (it is easy to see that the tensor product of canonical tensors is itself a canonical tensor). This basis is likely not an optimal basis for the projected state, and thus the truncation performed with respect to it is not optimal.

\begin{figure}[t]
  \centering
  \includegraphics[width=\columnwidth]{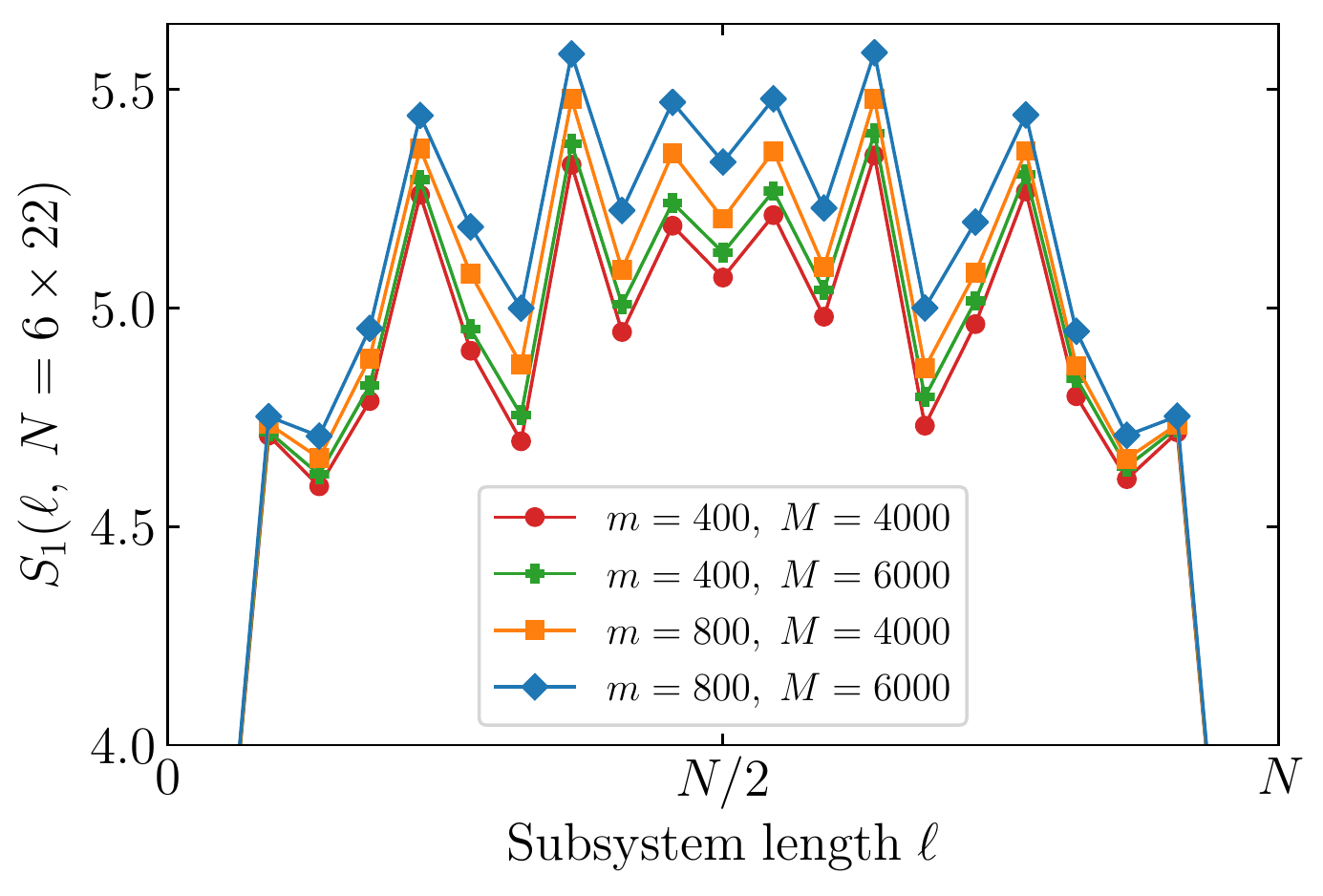}
  \caption{Entanglement entropy of the GPFS state obtained from a mean-field Hamiltonian with isotropic hopping on a 6-leg triangular lattice of size $6 \times 22$ with cylindrical boundary conditions for $m=400,800$ and $M=4000,6000$.
  \label{fig:6leghopvne}
  }
\end{figure}

In principle, this could be remedied by first constructing the product of the two parton MPS and applying the Gutzwiller projection without truncation, then bringing the resulting exact GPFS into canonical form, and then performing the truncation (either via the SVD, or an alternating least-squares procedure, where one iterates over each tensor in the trial GPFS and attempts to maximize the overlap with the exact GPFS; in this latter case, the exact GPFS need not be constructed explicitly). However, both of these methods are much more costly in terms of scaling with the bond dimension. Therefore, in practice, we find the Gutzwiller zipper method to be far more accurate, as much higher bond dimensions can be reached.

It is worth noting that the same issue in principle applies to many popular MPS methods, including the TEBD time evolution method~\cite{vidal2004efficient}. However, in most of those applications, the truncation is small (for example, since one performs a time evolution only over some very short timestep), and thus this issue has very little effect. In the Gutzwiller projection, on the other hand, the truncation could be very large, and thus the effect is more significant.

In practice, the accuracy of the approach is controlled by to the bond dimension $m$ of each parton state and the bond dimension $M$ of the Gutzwiller-projected state. The interplay between the two is shown in Fig.~\ref{fig:6leghopvne}, which shows the entanglement entropy of the GPFS MPS for a 6-leg system at $m=400,800$ and $M=4000, 6000$. Naturally, the best accuracy is obtained for both $m$ and $M$ maximal. In this particular case, it turns out that the second-best result is obtained for $m=800$ and $M=4000$, which is slightly more accurate than $m=400$ and $M=6000$. However, we have been unable to find a general rule for determining the best parameters; instead, the convergence of the desired physical quantity has to be checked against both $m$ and $M$

\section{Simulation details}

In this section, we specify several details pertinent to the numerical results presented in the main text.

\subsection{Triangular lattice clusters and $J$-$K$ ring-exchange model}

\begin{figure}[t]
  \centering
  \includegraphics[width=\columnwidth]{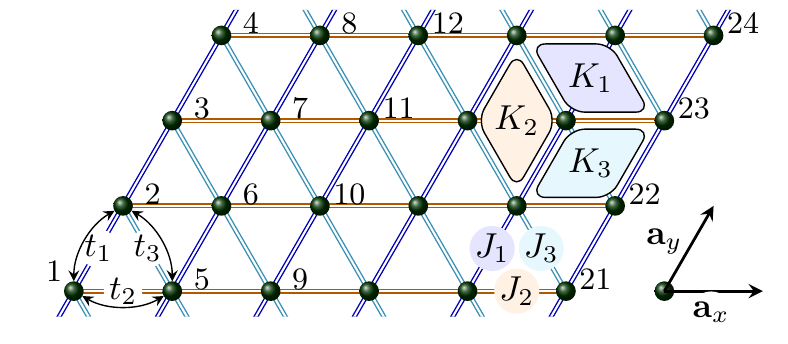}
  \caption{The type of triangular lattice clusters we consider, here drawn for an $N=L_y\times L_x = 4\times 6$ system with periodic boundary conditions in the $y$ direction. The chosen wrapping corresponds to the XC4 cylinder~\cite{yan_spin-liquid_2011}; e.g., site 1 is coupled to sites 4 and 8, site 5 is coupled to sites 8 and 12, etc. We also show a schematic representation of the $J$-$K$ spin Hamiltonianon (right) and mean-field parton Hamiltonian (bottom left). The site numbering specifies the DMRG path and thus the meaning of $\ell$ in all calculations of $S_1(\ell, N = L_y \times L_x)$.
  \label{fig:triangularlattice}
  }
\end{figure}

The family of $L_y$-leg triangular lattice clusters that we consider is depicted in Fig.~\ref{fig:triangularlattice}. On these lattice clusters, we simulate the SU(2) invariant Heisenberg antiferromagnet augmented by the four-site cyclic ring-exchange term introduced in the main text. The latter term performs a cyclic permutation of the spin configuration around a given four-site plaquette: $P_{ijkl}\ket{\sigma_i\sigma_j\sigma_k\sigma_l}=\ket{\sigma_l\sigma_i\sigma_j\sigma_k}$. The full Hamiltonian reads (following the conventions of Refs.~\cite{motrunich_variational_2005, sheng_spin_2009, block_spin_2011, mishmash_theory_2013}):
\begin{equation}
  \label{eq:spin-hamiltonian}
  H_{\text{spin}} = \sum_{\langle i,j\rangle} 2J_{ij}~\mathbf{S}_i\cdot \mathbf{S}_j
  + \sum_{ijkl \in \Diamond}
  K_{\Diamond}~(P_{ijkl} + \text{H.c.}).
\end{equation}
On ladder geometries, it is natural to allow anisotropic couplings as shown on the right side of Fig.~\ref{fig:triangularlattice}. The partons are described by a free fermion hopping Hamiltonian on the same lattice ($\uparrow$ and $\downarrow$ are assumed to have the same mean-field dynamics):
\begin{equation}
    \label{eq:parton-hamiltonian}
    H_{\text{MF}} =  - \sum_{\langle i, j\rangle} t_{ij}~c^{\dagger}_i c_j + \mathrm{H.c.},
\end{equation}
with hopping parameters also depicted in Fig.~\ref{fig:triangularlattice} (bottom left).

In the case of the 2-leg triangular strip~\cite{sheng_spin_2009}, we assume $J_1=J_3$ and $K_1=K_3=K$ in the spin model ($K_2$ plaquettes are absent) and $t_1=t_3$ in the spinon hopping Hamiltonian. When viewing the triangular strip as a 1D chain, $J_1$ and $t_1$ ($J_2$ and $t_2$) correspond to nearest-neighbor (next-nearest-neighbor) terms in the respective models. For $t_2/t_1 > 0.5$, the mean-field Hamiltonian emits a 2-band state ($N_\mathrm{slices}=2$); see Ref.~\cite{sheng_spin_2009} for all details.

\begin{figure}[b]
  \centering
  \includegraphics[width=0.49\columnwidth]{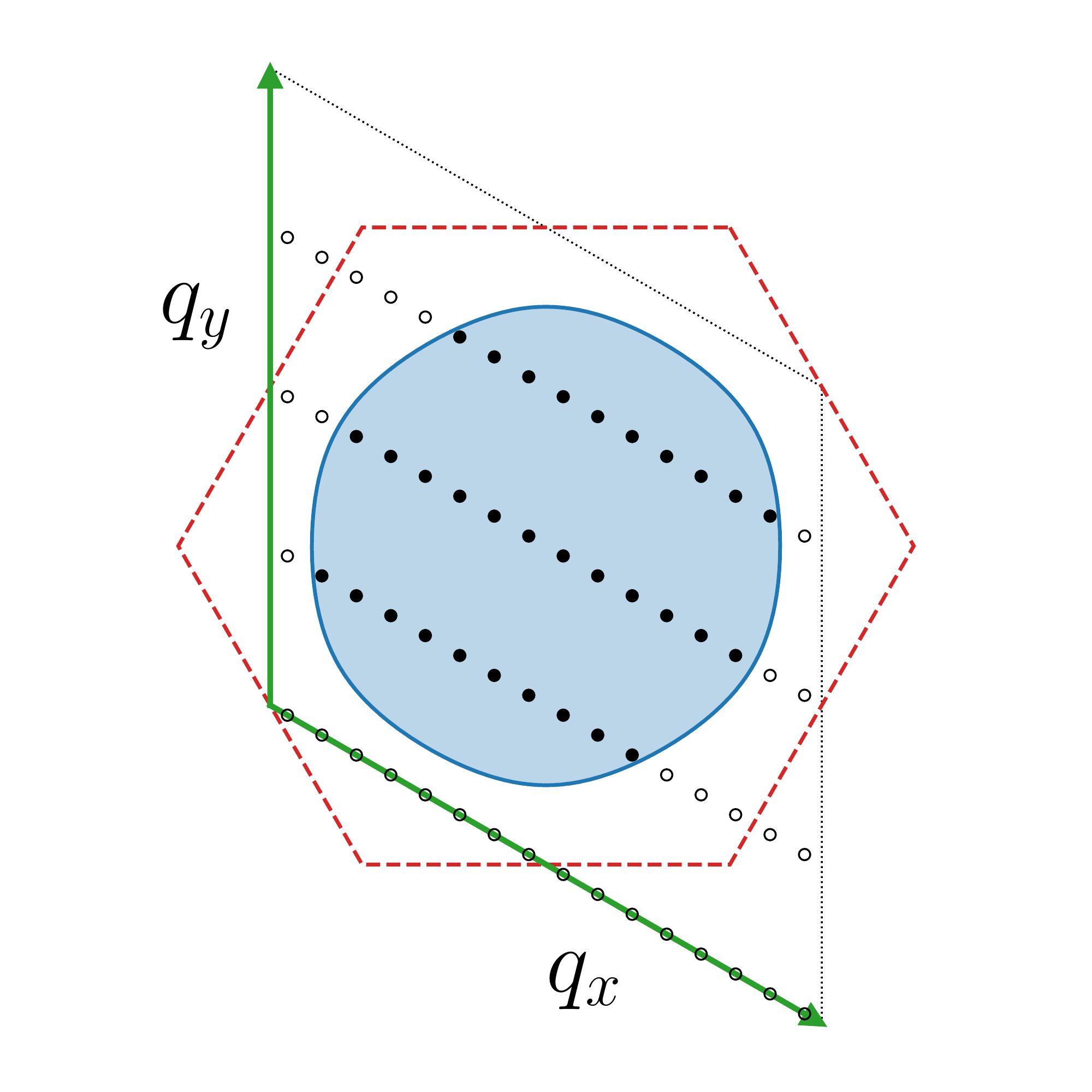}
  \includegraphics[width=0.49\columnwidth]{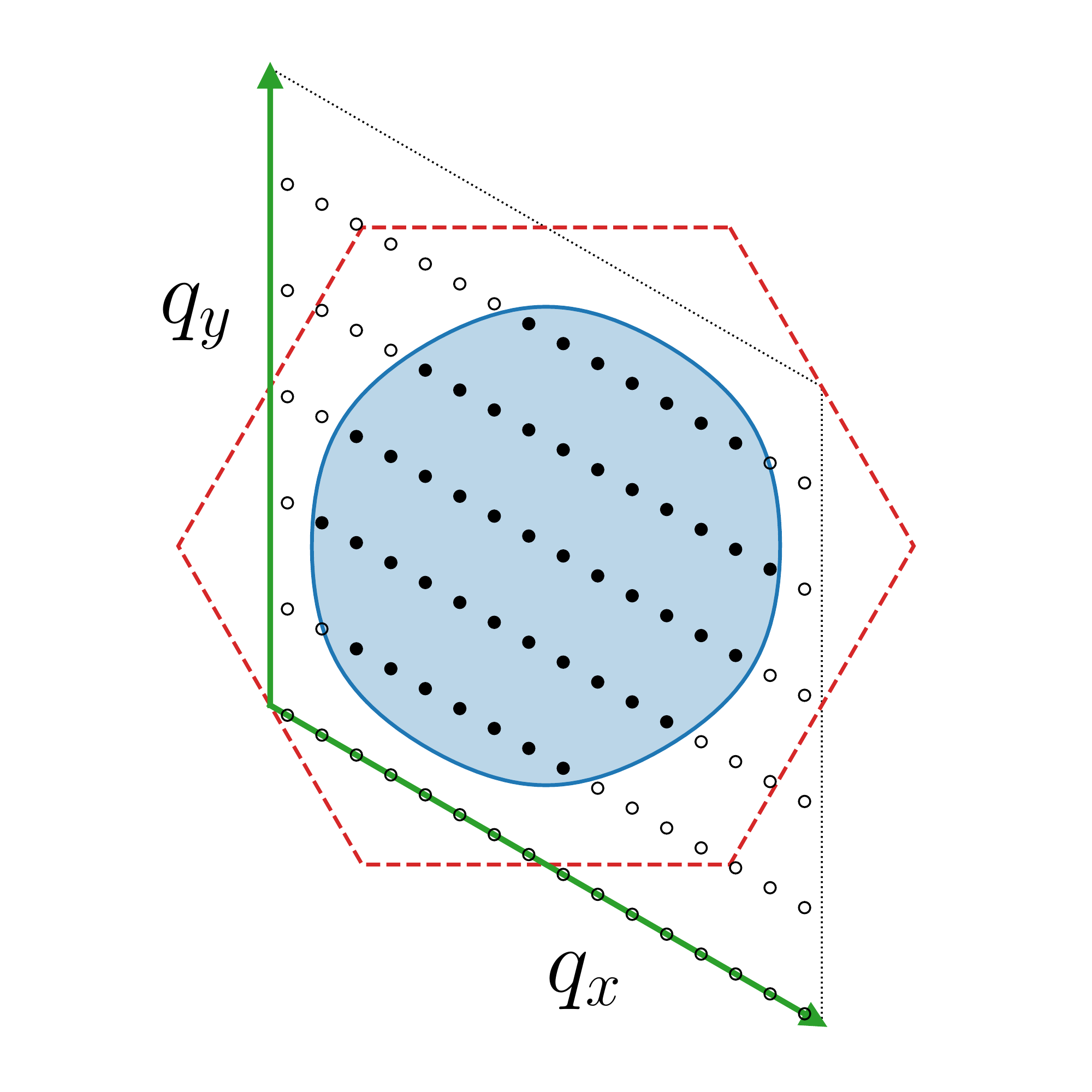}
  \caption{The Fermi sea of the 4-leg (left) and 6-leg (right) triangular ladders. The boundary condition is taken to be periodic in the $y$ direction and antiperiodic in the $x$ direction. The Fermi sea in the thermodynamic limit is depicted in blue (the Fermi surface is nearly circular at half filling on the triangular lattice). Illustrations of the filled orbitals are shown for system sizes of $N=4\times 16$ and $N=6\times 16$, respectively.
  \label{fig:triangular4,6-BZ}
  }
\end{figure}

For our studies of the 4- and 6-leg systems, we take isotropic couplings $J_i=J$ and $K_i=K$; similarly we only consider isotropic mean-field hopping patterns with $t_i=t$ when defining the GPFS trial states. The Brillouin zone for the triangular lattice with the allowed discrete momenta (for a toroidal system) on 4- and 6-leg ladders is shown in Fig.~\ref{fig:triangular4,6-BZ}. The 4-leg (6-leg) states have 3 (5) bands, i.e., $N_\mathrm{slices}=3$ (5) cuts through the Fermi surface (see also, e.g., Refs.~\cite{geraedts_half-filled_2016, mishmash_entanglement_2016, szasz_chiral_2020}). The central charge of the mean-field state is $c = 2N_\mathrm{slices}$ (with 2 due to spin), while the corresponding GPFS will have $c = 2N_\mathrm{slices} - 1$, as Gutzwiller projection will remove the overall conducting charge mode.

\subsection{Entanglement entropy: Definitions and fitting}
\label{sec:EEdetails}

Given the reduced density matrix $\rho_A$ for some subset of the system, the Renyi entanglement entropy is given by
\begin{equation}
    S_\alpha(\rho_A) = \frac{1}{1 - \alpha} \log\left(\mathrm{Tr}\,\rho_A^\alpha \right),
\end{equation}
where $\alpha$ is the Renyi index. For $\alpha=1$, the conventional von Neumann entanglement entropy is recovered:
\begin{equation}
    S_1(\rho_A) = -\operatorname{Tr} \left[ \rho_A \log(\rho_A) \right].
\end{equation}
To extract the central charge from the entanglement entropy for a subregion of the $\ell$ leftmost sites (see Fig.~\ref{fig:triangularlattice}), we use the formula of Calabrese and Cardy~\cite{0405152}:
\begin{equation}
  \label{eq:calabrese-cardy-scaling}
  S_1(\ell, N = L_y\times L_x) = \frac{c}{6}\log{\left(\frac{N}{\pi}\sin
      \frac{\ell\pi}{N}\right)} + A',
\end{equation}
which assumes open boundary conditions in the long ($x$) direction (it would be twice that for systems with periodic boundary conditions). 

We use the \texttt{curve\_fit} function from the \texttt{scipy.optimize} package~\cite{scipy} to fit Eq.~\eqref{eq:calabrese-cardy-scaling} to the entanglement entropy of the GPFS MPS obtained from the Gutzwiller zipper and the final DMRG ground states. For the 4- and 6-leg ladders, only the subsystems corresponding to a full cut through cylinder were included in the fits, i.e., clean ``rung cuts'' corresponding to every 4 / 6 sites. The inclusion of small subsystems, especially on systems with open boundary conditions, has a significant impact on the obtained fit value for the central charge. Since this data near the edge of the sample is strongly polluted by nonuiversal boundary effects, it is best to exclude some portion of the data from the two sides when performing the fits. For example, in the $2 \times 48$ system of Fig.~\ref{fig:2legcomparison}, we excluded eight sites from each side, while for the 4-leg data in Fig.~\ref{fig:4leggutzvne}, we excluded seven rungs. For the fully periodic boundary conditions data presented below in Fig.~\ref{fig:4-legs-fullperiodic}, two rungs have been excluded. We note that the most robust way to estimate the central charge on such finite systems is to analyze the scaling of the entanglement entropy for $\ell=N/2$ (i.e., the half-system entanglement entropy cut) versus $L_x$, as shown in the inset of Fig.~\ref{fig:4leggutzvne}.

\begin{figure*}[t]
  \centering
  \includegraphics[width=0.95\textwidth]{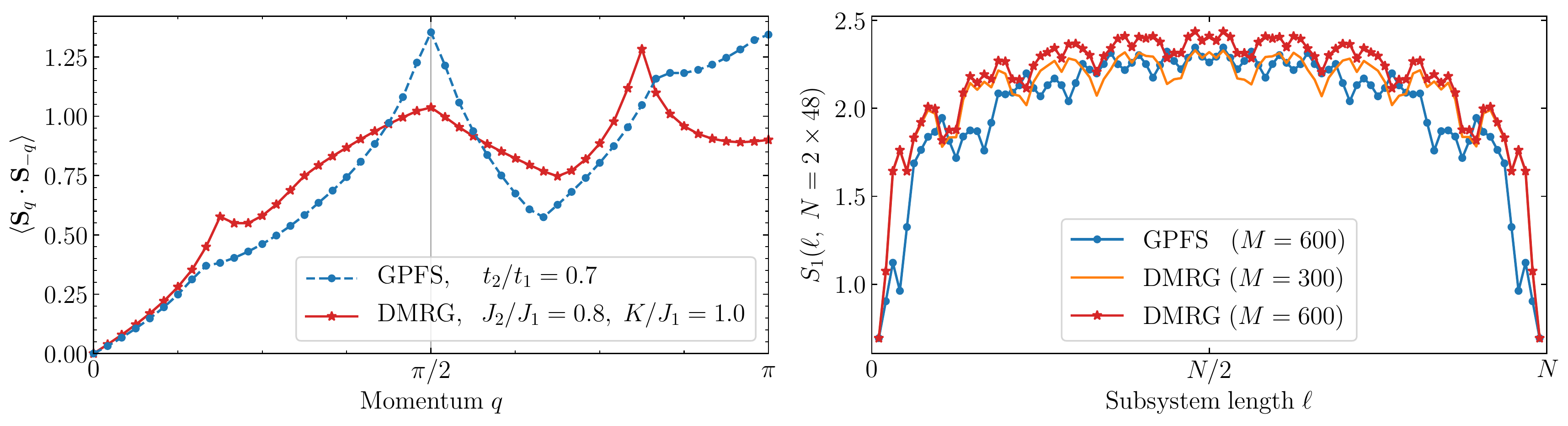}
  \includegraphics[width=0.95\textwidth]{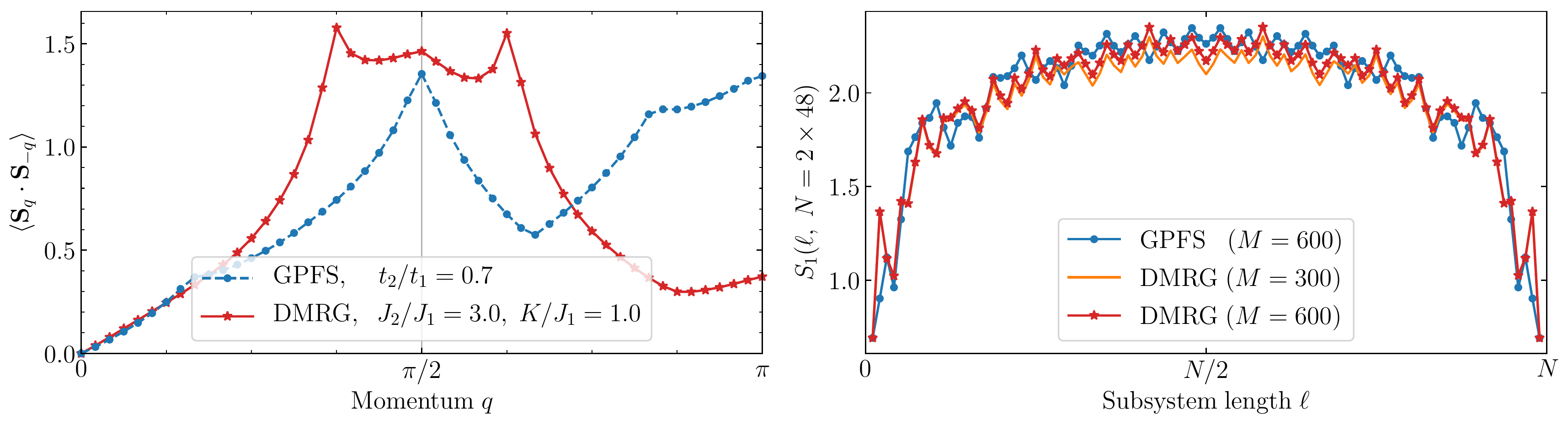}
  \includegraphics[width=0.95\textwidth]{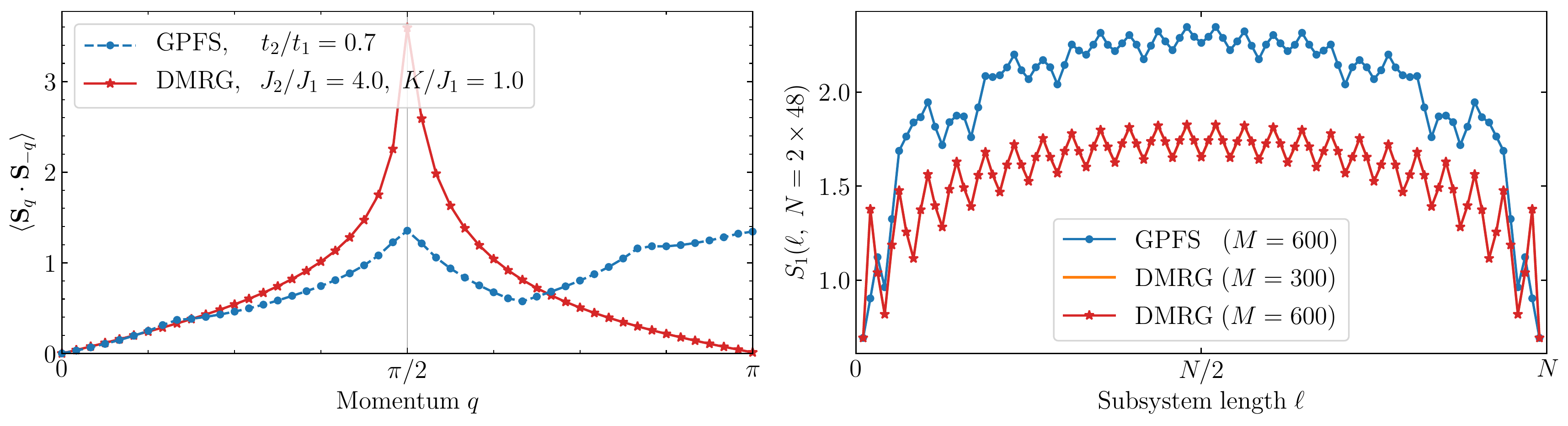}
  \includegraphics[width=0.95\textwidth]{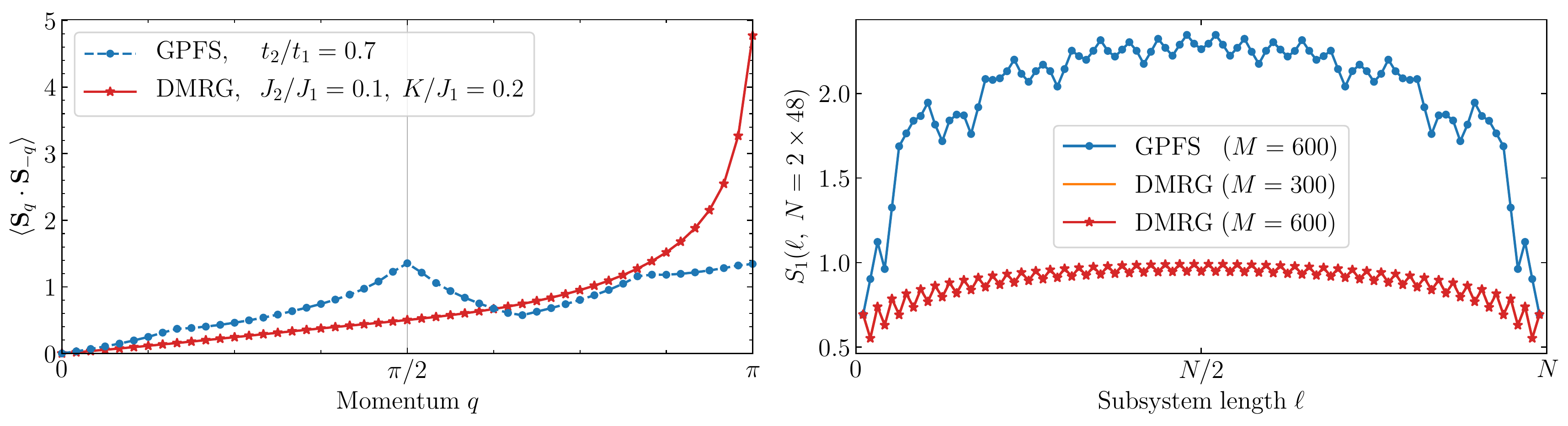}
  \caption{Exploring the phase diagram of the 2-leg ladder $J_1$-$J_2$-$K$ model by applying two sweeps of DMRG on a single 2-band GPFS state. The spin structure factors are shown on the left, while von Neuman entanglement entropy data is shown on the right. All rows have $K/J_1=1$. The first row $J_2/J_1=0.8$ corresponds to the large wave vector SFS, %the second row $J_2/J_1=2$ correspond to VBS-3state 
  the second row $J_2/J_1=3$ corresponds to the small wave vector SFS phase between VBS-2 and VBS-3 in Fig.~2 of Ref.~\cite{sheng_spin_2009}, the third row $J_2/J_1=4$ corresponds to the VBS-2 phase, and the last row $J_2/J_1=0.1$ to the Bethe chain phase. Indeed a drop in the entanglement entropy dome occurs when the realized phase has a lower central charge. Also, note that the initial GPFS state (blue data) is chosen to be the same for all choices of DMRG parameters. For all GPFS MPS data, we have taken $m=200$ and $M=600$; the $M=300$ DMRG runs in the right panels were initialized with a GPFS state with $m=200$ and $M=300$. Note that in the inset of Fig.~\ref{fig:2legcomparison} of the main text, the energy of the DMRG (with GPFS initialization) slightly increases (comparable to the square root of machine precision) after the fourth sweep and before reaching its final converged value; we have checked that those additional sweeps cause no noticeable change in physical quantities.
  \label{fig:2legs}
  }
\end{figure*}

\subsection{Spin structure factor}

The spin structure factor we compute is defined as
\begin{equation}
  \label{eq:sqsq}
  \langle \mathbf{S}_\mathbf{q}\cdot\mathbf{S}_{-\mathbf{q}} \rangle = \frac{1}{N}\sum_{\mathbf{r}, \mathbf{r}'}\mathrm{e}^{-\mathrm{i}\mathbf{q}\cdot(\mathbf{r}-\mathbf{r}')}
  \langle \mathbf{S}_{\mathbf{r}}\cdot\mathbf{S}_{\mathbf{r}'} \rangle.
\end{equation}
Although we mainly work on open cylinders, we still use this form for the structure factor as the averaging over different ``origins'' serves to effectively wash out effects of boundary-condition-induced breaking of translational symmetry.

In Eq.~\eqref{eq:sqsq}, the site positions are $\mathbf{r}=n_y\mathbf{a_y} + n_x \mathbf{a_x}$, where $\mathbf{a}_{y}, \mathbf{a}_{x}$ are the primitive translation vectors of the triangular lattice (see Fig.~\ref{fig:triangularlattice}) with $n_y=0, \dots, L_y-1$ and $n_x = 0, \dots, L_x-1$. $\mathbf{q}=(q_y, q_x)$ is a reciprocal lattice vector (see Fig.~\ref{fig:triangular4,6-BZ}). As our systems are narrow and periodic in the $y$ direction, we have quantized momenta
\begin{equation}
    q_y = m_y\frac{2\pi}{L_y}, \quad m_y=0, \dots, L_y-1.
\end{equation}
In Figs.~\ref{fig:4legdmrggutz}, and \ref{fig:6leg-sqsq}, the structure factors are plotted for each $q_y$ separately as a function $q_x$. (We  plot only values of longitudinal momenta $q_x = m_x\frac{2\pi}{L_x}$ with $m_x=0, \dots, L_x-1$, although on an open cylinder this is not required.) In Fig.~\ref{fig:2legs}, we treat the 2-leg triangular strip as a 1D chain such that $\mathbf{q}\to q$ is a 1D momentum~\cite{sheng_spin_2009}.

\section{Additional supporting data}

In this section, we present additional data benchmarking our approach on the 2-leg ladder and filling in various details of the situation on the 4- and 6-leg ladders.

\subsection{Benchmarking ``warm starting'' DMRG with GPFS MPS on the 2-leg triangular strip ring model}

In this section, we use the $J_1$-$J_2$-$K$ ring model on the 2-leg triangular strip~\cite{sheng_spin_2009} as a testbed to benchmark the GPFS state preparation strategy used throughout. We choose a generic 2-band GPFS ansatz with fixed $t_2/t_1$ and construct the corresponding MPS via the Gutzwiller zipper. Using this highly entangled ``mother'' state to warm start DMRG, we subsequently run $\mathcal{O}(1)$ DMRG sweeps for various points in the phase diagram ($J_2/J_1$, $K/J_1$). Rather remarkably, merely two DMRG sweeps is able to accurately reproduce the entire phase diagram of the model, including within phases markedly distinct from the starting 2-band GPFS state.

\begin{figure}[b]
  \centering
  \includegraphics[width=\columnwidth]{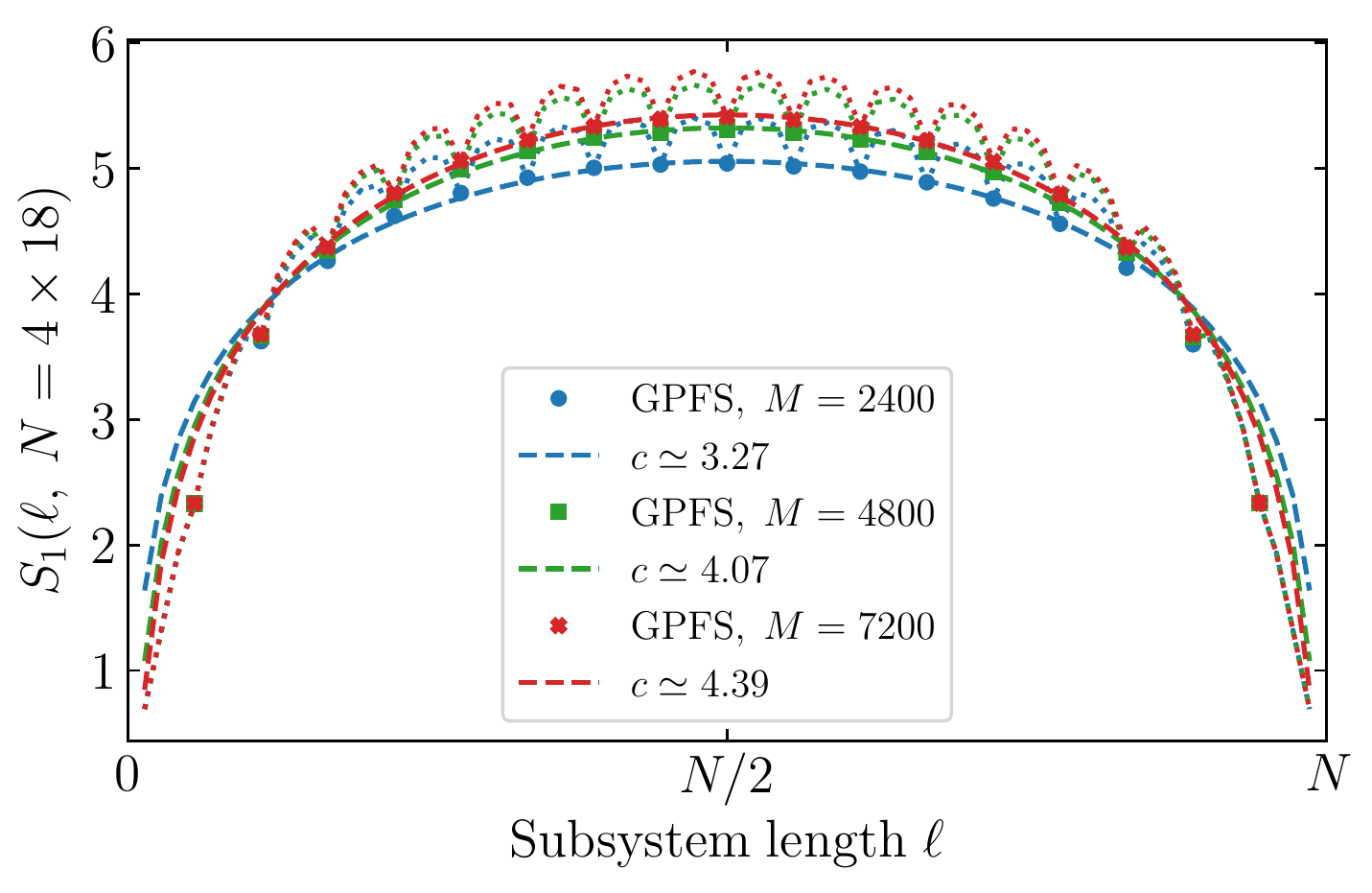}
  \caption{Entanglement entropy $S_1$ as a function of subregion size for the $N = 4\times18$ 4-leg triangular ladder taking \emph{fully periodic} boundary conditions for the spin wave function [for the partons, we take periodic (antiperiodic) boundary conditions in $y$ ($x$)]. The fits were performed on subregions corresponding to straight cuts through the cylinder ring (here every four sites; points connected by dashed lines) and two points are excluded from each side.
  \label{fig:4-legs-fullperiodic}
  }
\end{figure}

In Fig.~\ref{fig:2legs}, we show spin structure factors (left panels) and entanglement entropy curves (right panels) for a GPFS MPS with $t_2/t_1=0.7$ as well as the final DMRG data taken at values of $J_2/J_1$ and $K/J_1$ indicated in the legends of the left panels; each row corresponds to a different set of spin Hamiltonian parameters. In all cases, the DMRG data is consistent with the phase quoted in Ref.~\cite{sheng_spin_2009} (with any quantitative difference attributable to differences in chosen boundary conditions). The first row is the same U(1) SFS point as in Fig.~\ref{fig:2legcomparison} (note the slight renormalization of the SFS going from the trial state to the final DMRG state). The second row is the ``large $J_2$, $t_2$'' SFS state of~\cite{sheng_spin_2009}; the realized state can be obtained from the starting trial state at $t_2/t_1=0.7$ via a drastic SFS renormalization, although the two phases are not continuously connected in the phase diagram of the $J_1$-$J_2$-$K$ model itself. In each case, in the right panels we see that the entanglement scaling remains nearly at a ``fixed point'' upon running DMRG on the trial state (modulo a slight increase in the constant piece $A'$). This occurs even when we purposefully decrease $M$ (e.g., to $M=300$); that is, even if the bond dimension is not large enough to \emph{fully} converge the final ground state, the entanglement entropy roughly stays put after DMRG. This gives us confidence that there is no pathological behavior upon running DMRG on top of a GPFS MPS initial state when the chosen $M$ is insufficient for full convergence, as happens on the 6-leg system. The third row is within the VBS-2 phase at large $J_2/J_1$, which is close to the decoupled chains limit of the model. Here the final state has a lower entanglement / central charge yet DMRG quickly finds the correct state, which is reminiscient of the behavior found on the 4- and 6-leg systems in Figs.~\ref{fig:4legdmrggutz} and \ref{fig:6legdmrggutz} of the main text. Finally, the bottom row of Fig.~\ref{fig:2legs} corresponds to the $c=1$ Bethe chain phase, which can be understood as a 1-band SFS. In other words, DMRG effectively completely renormalizes away the smaller Fermi pocket in the seed GPFS state. Again, the true ground state has a lower central charge ($c=1$) than the starting state ($c=3$), yet the former is efficiently found by the DMRG.

\subsection{GPFS MPS with fully periodic boundary conditions on the 4-leg ladder}

For the 4-leg ladder, we have also used the Gutzwiller zipper to obtain the GPFS MPS with \emph{periodic} boundary conditions as employed by Block et al.~\cite{block_spin_2011}. As shown in Fig.~\ref{fig:4-legs-fullperiodic}, despite taking a bond dimension up to $M=7200$, the entanglement entropy of the trial state does not completely converge, although a clean dome clearly forms and the data is trending toward eventual $c=5$ scaling.

Guided by Fig.~\ref{fig:4legdmrggutz} of the main text, we would expect the constant piece of the entanglement entropy for the final putative $c=0$ state of the $J$-$K$ model (at $K/J=0.6$) to be $S_1 \approx 5.5~(=2\cdot2.75)$ on this fully periodic system. Thus, even the $N=4\times18$ PBC cluster of Fig.~\ref{fig:4-legs-fullperiodic} is likely too small of a system to obtain conclusive results on this model using our strategy.

\subsection{$S_1$ convergence and structure factor data on the 6-leg ladder}

\begin{figure}[t]
  \centering
  \includegraphics[width=\columnwidth]{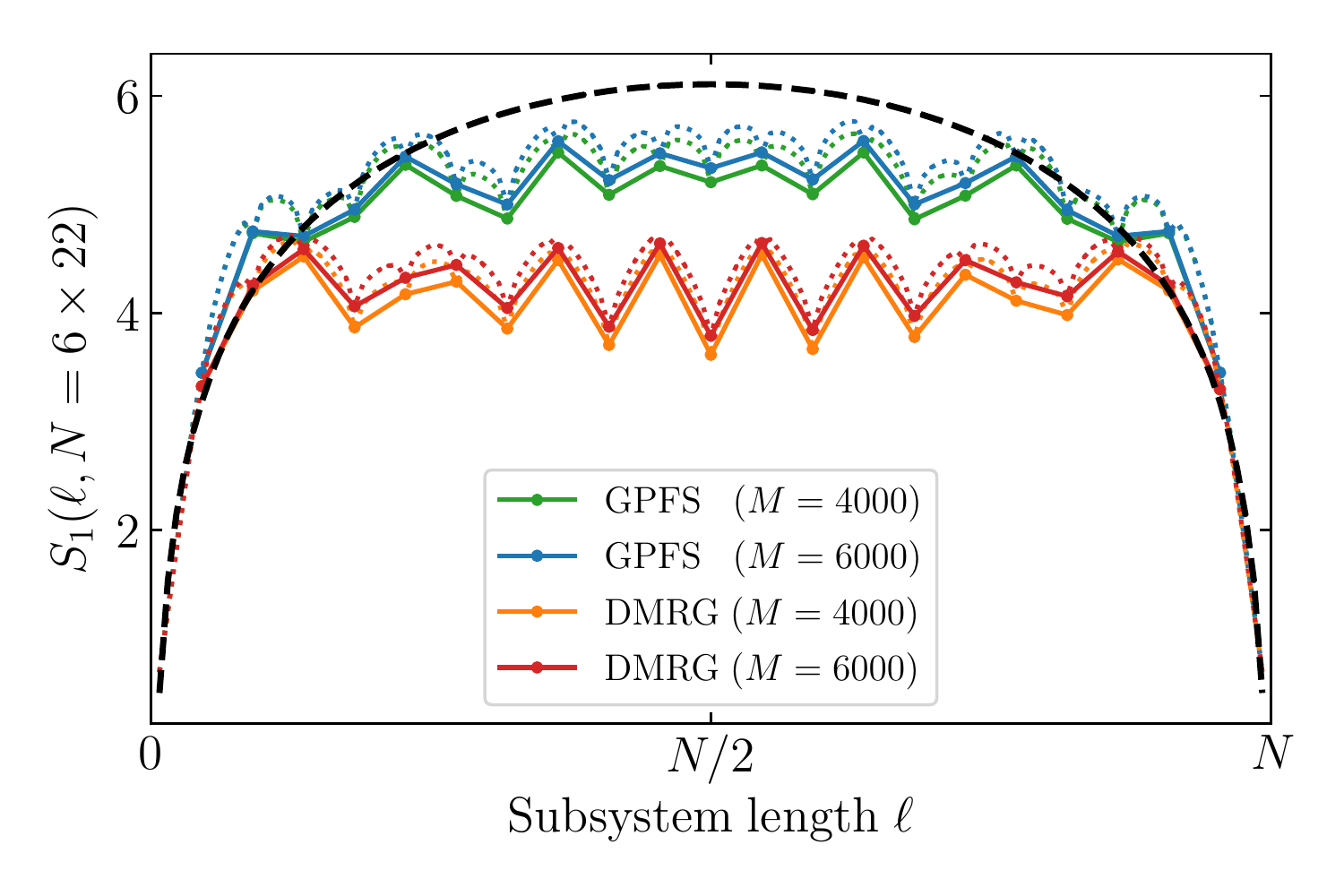}
  \caption{The von Neumann entanglement entropy as a function of subsystem size for the 6-leg triangular ladder on an $N = 6\times 22$ cluster with cylindrical boundary conditions (XC6 cylinders) for both the GPFS MPS and final DMRG states (cf.~Fig.~\ref{fig:6legdmrggutz} of the main text). We show data for both $M=4000$ and 6000 to illustrate the (rather weak) dependence on $M$; the bond dimension of the parton MPSs is $m=800$. For reference, the dashed curve corresponds to true $c=9$ scaling with (an arbitrarily chosen) $A'=0.5$; this is approximately the  entropy we would expect for a fully converged GPFS MPS.
  \label{fig:6-legs}
  }
\end{figure}

In Fig.~\ref{fig:6-legs}, we show the 6-leg GPFS MPS and final DMRG state at $K/J=0.6$ with different bond dimensions $M=4000,6000$ (cf.~Fig.~\ref{fig:6legdmrggutz} of the main text). While the final DMRG entanglement entropy is still not fully converged in $M$, the fact that we are starting the DMRG optimization in a state of \emph{higher} entanglement makes an eventual $c=9$ (or 8) state seem unlikely (cf.~Fig.~S11 of Ref.~\cite{he_spinon_2018}; note that the parameters in that figure correspond to $K/J=\infty$ in our model---see Sec.~VIII of the Supplemental Material of Ref.~\cite{he_spinon_2018} for a translation of conventions). That is, we know that we can capture entanglement entropy values of nearly $S_1 \sim 6$ with these bond dimensions (i.e., the GPFS MPS); however, the DMRG clearly prefers a lower entanglement ground state. We believe the final DMRG entropy values near $\ell\sim N/2$ are nearly converged (although the nature of the strong rung-to-rung oscillation is a feature of the data to be understood in future work). We would also like to point out that focusing only on the entanglement entropy values near the edge of the sample is clearly problematic---the dashed curve in Fig.~\ref{fig:6-legs} corresponds to $c=9$ (with $A'=0.5$), and while it tracks the DMRG entanglement for the first rung or two, this is unlikely very meaningful  in light of the arguments made herein.

\begin{figure}[t]
  \centering
  \includegraphics[width=\columnwidth]{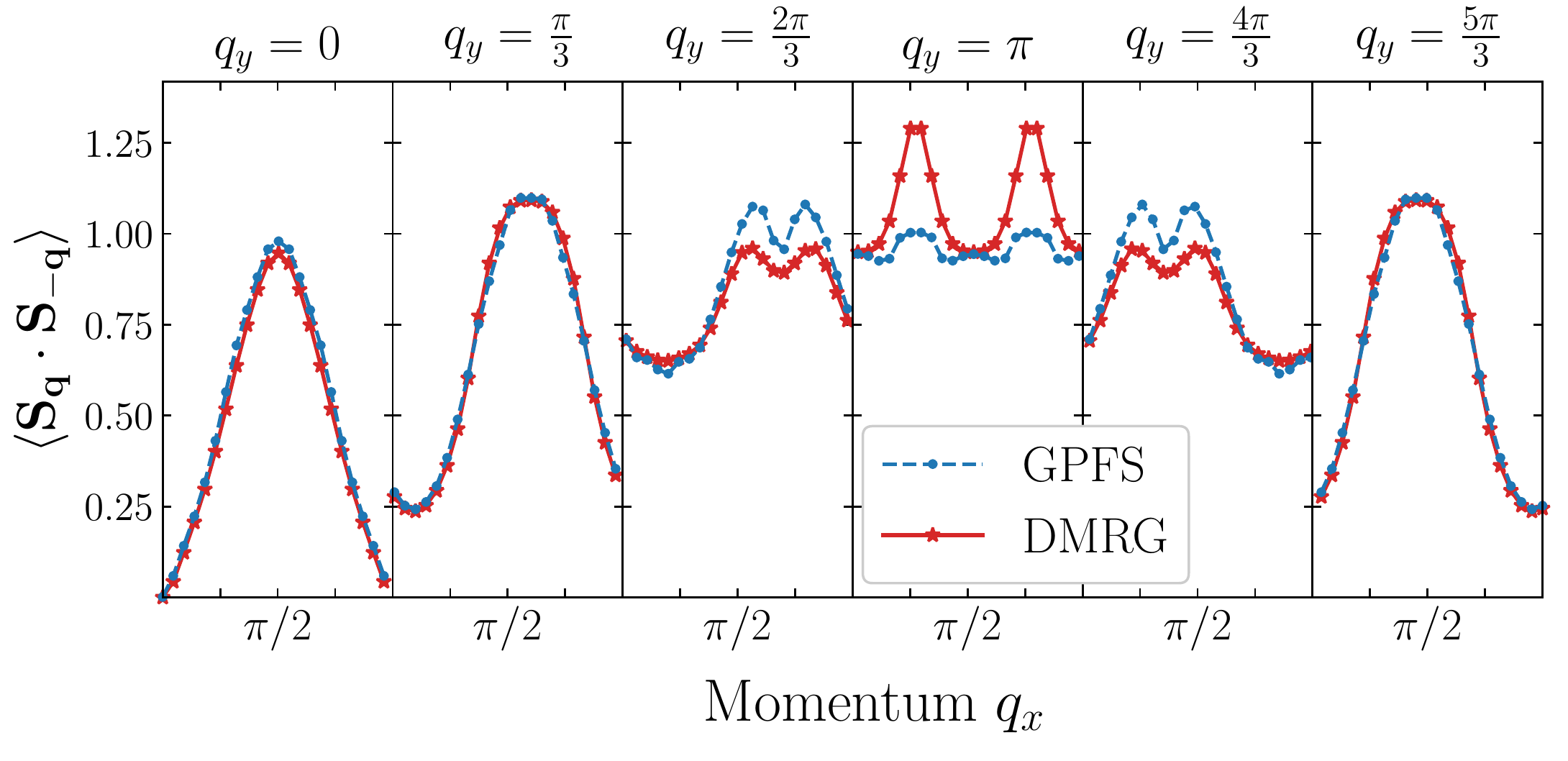}
  \includegraphics[width=\columnwidth]{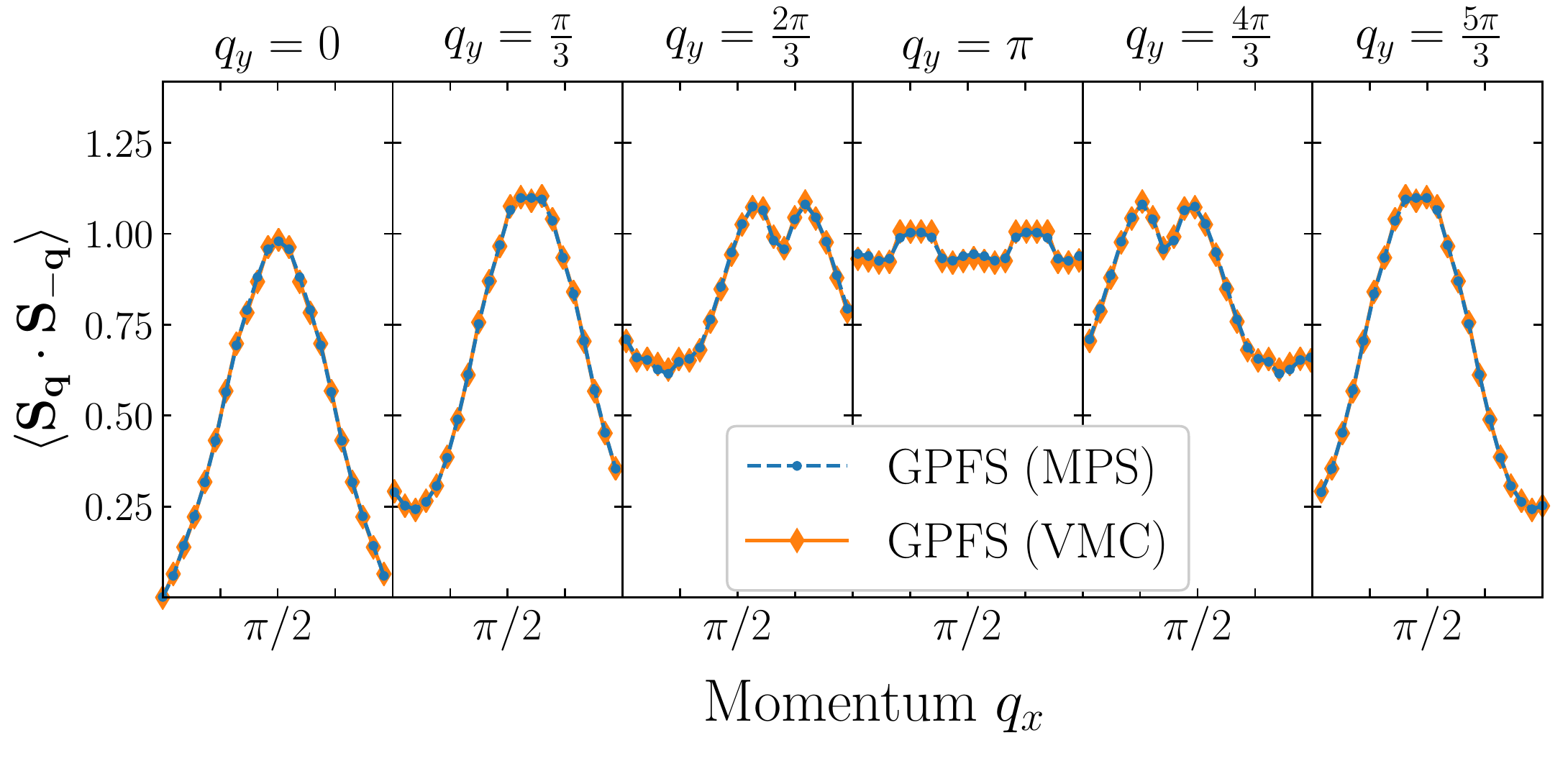}
  \caption{Top: Spin structure factors for the GPFS MPS and final DMRG states on the 6-leg triangular ladder (cf.~the corresponding 4-leg data in the bottom panel of Fig.~\ref{fig:4legdmrggutz} in the main text). Bottom: The spin structure factor of the GPFS state obtained as an MPS via the Gutziller zipper [GPFS (MPS)] and via standard VMC sampling [GPFS (VMC)]~\cite{gros_physics_1989}. Truncation error in the former case thus appears to give only negligible error in the structure factor, although the wave function's entanglement entropy is not quite fully converged (see Fig.~\ref{fig:6-legs}).
  \label{fig:6leg-sqsq}
  }
\end{figure}

We present in Fig.~\ref{fig:6leg-sqsq} spin structure factor data for the 6-leg ladder. We show in the top panel of Fig.~\ref{fig:6leg-sqsq} a plot analogous to the 4-leg data presented in the bottom panel of Fig.~\ref{fig:4legdmrggutz} of the main text, comparing the GPFS MPS and final DMRG structure factors (at $K/J=0.6$). The qualitative agreement of $\langle \mathbf{S}_\mathbf{q} \cdot \mathbf{S}_{-\mathbf{q}}\rangle$ between the two states is quite good, and some features even seem to get ``enhanced'' upon running DMRG. However, as stressed above, the entanglement scaling of the DMRG state is nearly flat (modulo oscillations). Indeed there are some hints of such gap formation in the DMRG structure factor data in Fig.~\ref{fig:6leg-sqsq}; e.g., the slight ``softening'' of $\langle \mathbf{S}_\mathbf{q} \cdot \mathbf{S}_{-\mathbf{q}}\rangle$ near $\mathbf{q}=0$ ~\cite{mishmash_continuous_2015} relative to the GPFS state.

Finally, in the bottom panel of Fig.~\ref{fig:6leg-sqsq}, we show measurements of $\langle \mathbf{S}_\mathbf{q} \cdot \mathbf{S}_{-\mathbf{q}}\rangle$ taken with respect to the GPFS MPS and the same trial state sampled via traditional VMC techniques. Indeed the two agree very well, indicating that the GPFS MPS is still accurately capturing some long-distance features of the state (e.g., power laws) even without fully converged entanglement. All in all, the 6-leg GPFS MPS seems well-behaved and has significant entanglement; thus, there is no obvious reason for DMRG to \emph{decrease} the entanglement, that is unless the U(1) SFS is not a correct description of the true ground state of the model.

% \newpage
% \bibliography{./refs}

\end{document}